\documentclass{article}

\usepackage{microtype}
\usepackage{graphicx}
\usepackage{subfigure}
\usepackage{booktabs} % for professional tables

\usepackage{hyperref}

\newcommand{\msESM}{\textsc{ESM-AA}}

\usepackage{multirow}
\usepackage[table]{xcolor}
\definecolor{mygray}{gray}{.9}
\usepackage[figuresright]{rotating}

\usepackage[accepted]{icml2024}

\usepackage{amsmath}
\usepackage{amssymb}
\usepackage{mathtools}
\usepackage{amsthm}

\usepackage[capitalize,noabbrev]{cleveref}

\theoremstyle{plain}

\theoremstyle{definition}

\theoremstyle{remark}

\usepackage[textsize=tiny]{todonotes}

\icmltitlerunning{ESM All-Atom: Multi-scale Protein Language Model for Unified Molecular Modeling}

\begin{document}

\twocolumn[
\icmltitle{ESM All-Atom: Multi-scale Protein Language Model for Unified Molecular Modeling}



\icmlsetsymbol{equal}{*}
\icmlsetsymbol{corresponding}{\#}

\begin{icmlauthorlist}
\icmlauthor{Kangjie Zheng}{equal,pku}
\icmlauthor{Siyu Long}{equal,nju}
\icmlauthor{Tianyu Lu}{thu}
\icmlauthor{Junwei Yang}{pku}\\
\icmlauthor{Xinyu Dai}{nju}
\icmlauthor{Ming Zhang}{corresponding,pku}
\icmlauthor{Zaiqing Nie}{air,PharMolix}
\icmlauthor{Wei-Ying Ma}{air}
\icmlauthor{Hao Zhou}{corresponding,air}
\end{icmlauthorlist}
\icmlaffiliation{pku}{School of Computer Science, National Key Laboratory for Multimedia Information Processing, Peking University-Anker Embodied AI Lab, Peking University, Beijing 100871, China}
\icmlaffiliation{nju}{School of Artificial Intelligence, National Key Laboratory for Novel Software Technology, Nanjing University}
\icmlaffiliation{thu}{Department of Computer Science, Tsinghua University}
\icmlaffiliation{air}{Institute for AI Industry Research (AIR), Tsinghua University. This work was done during the internship of Kangjie, Siyu, Tianyu, and Junwei at AIR}
\icmlaffiliation{PharMolix}{PharMolix Inc}


\icmlcorrespondingauthor{Hao Zhou}{zhouhao@air.tsinghua.edu.cn}
\icmlcorrespondingauthor{Ming Zhang}{mzhang\_cs@pku.edu.cn}
\icmlkeywords{Machine Learning, ICML}

\vskip 0.3in
]




\printAffiliationsAndNotice{\icmlEqualContribution} % otherwise use the standard text.

% Our experimental results demonstrate that \msESM~outperforms previous methods in protein-molecule tasks.
% Further analysis reveals that \msESM~outperforms many molecule-specific models on standard molecular benchmarks without compromising protein understanding.

\begin{abstract}
Protein language models have demonstrated significant potential in the field of protein engineering.
However, current protein language models primarily operate at the residue scale, which limits their ability to provide information at the atom level.
This limitation prevents us from fully exploiting the capabilities of protein language models for applications involving both proteins and small molecules.
In this paper, we propose \msESM~(ESM All-Atom), a novel approach that enables atom-scale and residue-scale unified molecular modeling.
\msESM~achieves this by pre-training on multi-scale code-switch protein sequences and utilizing a multi-scale position encoding to capture relationships among residues and atoms.
Experimental results indicate that \msESM~surpasses previous methods in protein-molecule tasks, demonstrating the full utilization of protein language models.
Further investigations reveal that through unified molecular modeling, \msESM~not only gains molecular knowledge but also retains its understanding of proteins. \footnote{The source codes of \msESM~ are publicly released at \href{https://github.com/zhengkangjie/ESM-AA}{https://github.com/zhengkangjie/ESM-AA}.}

\end{abstract}

\section{Introduction}
\label{introduction}
Protein language models (PLMs) have demonstrated significant potential in protein engineering, enabling the capture of biochemical and co-evolutionary knowledge during the pre-training of large-scale protein sequences.
This has resulted in remarkable achievements across various domains, including protein structure prediction \citep{wu2022high,fang2022helixfold_single}, protein fitness prediction \citep{mardikoraem2023protein,notin2022tranception}, protein design \citep{zheng2023structure,ferruz2022protgpt2}, etc.
For instance, ESM \citep{rives2021biological,lin2022language}, a widely used PLM, has served as the foundation for several significant models, including ESM-Fold \citep{lin2023evolutionary} for precise protein structure prediction and LM-Design \citep{verkuil2022language,hie2022high} for designing proteins with given target functions.

Current PLMs primarily operate at the \textit{protein residue} (amino acid) \textit{scale}, which does not provide information at the \textit{atom scale}.
In such circumstances, the potential of PLMs cannot be fully exploited to benefit applications involving both macromolecules (proteins) and small molecules, both of which are vital for various downstream applications.\footnote{These applications are widespread in the fields of chemistry and biology and are consistently pivotal for specific scientific breakthroughs. For instance, drug discovery aims to identify small molecules capable of binding to protein pockets \citep{anderson2003process,batool2019structure}, while enzyme engineering seeks to find enzymes (a special type of protein) that can efficiently catalyze molecular reactions \citep{mazurenko2019machine,kroll2023general}.} 
Thus, external small molecule models must be included to address these applications.
However, proteins are also composed of atoms, and modeling proteins solely at the residue scale may result in low resolution, meaning that it might not capture information at the atom scale.
Intuitively, extending PLMs to operate at both residue and atom scales would make them applicable to a larger range of applications.

\begin{figure*}[!t]
% \vspace{-0.8cm} 
\centering
\includegraphics[scale=0.34]{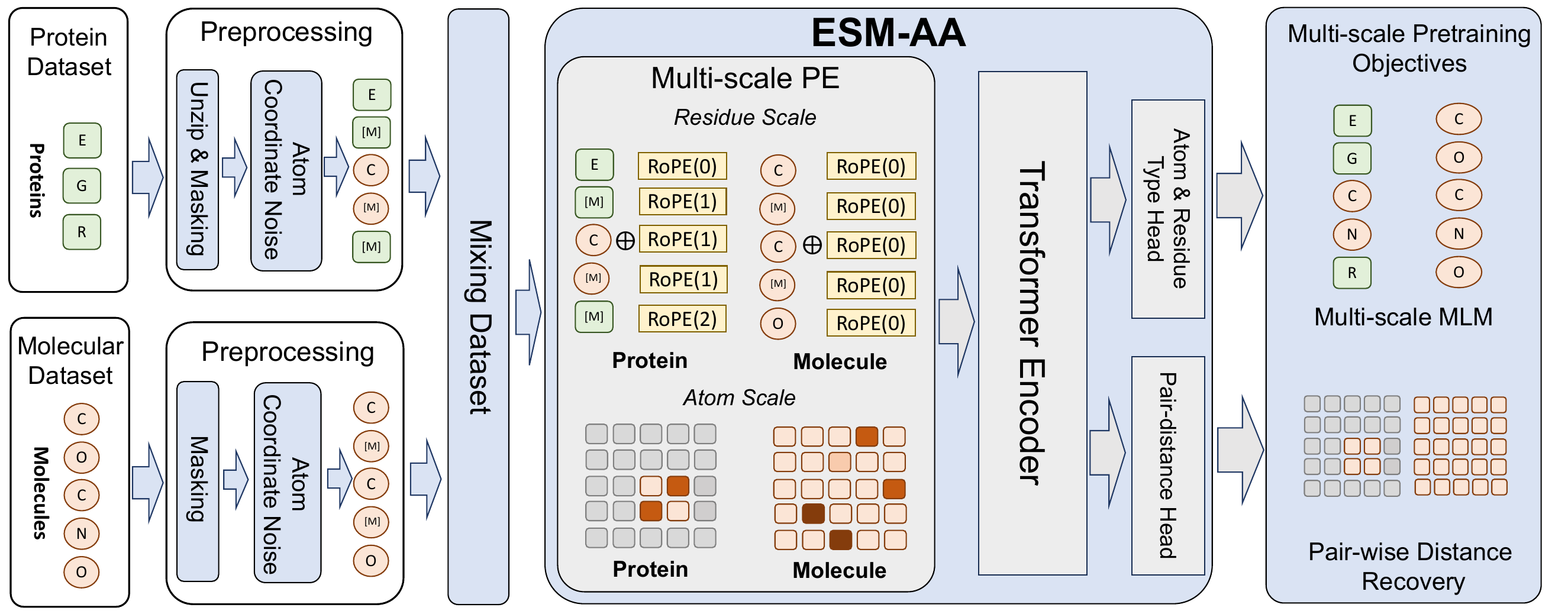}
\caption{Overview of our multi-scale pre-training process. We mix protein datasets and molecular datasets to train \msESM. It is worth noting that the model's input is either a molecule or a protein, but not paired protein-molecule data.}
\label{fig_overview}
\vspace{-0.3cm} 
\end{figure*}

Nevertheless, the development of multi-scale PLMs poses significant challenges.
First, achieving \textit{unified molecular modeling} that operates effectively at both the residue and atom scales is a challenging task, due to the incompatible vocabularies used at these two different scales.
One potential approach to incorporate atomic information into PLMs is to represent and pre-train proteins at the atom scale instead of the original residue-scale pre-training.
However, it should be noted that a typical protein can consist of thousands of residues, containing hundreds of thousands of atoms, making such an approach inefficient for modeling.
Second, designing an appropriate position encoding to accurately describe the relationships among residues and atoms within the same protein is also non-trivial, which involves relationships varying from residues to residues, residues to atoms, and atoms to atoms.

To tackle the aforementioned challenges, in this paper, we propose \msESM~(ESM All-Atom), which achieves multi-scale unified molecular modeling through (\romannumeral1) pre-training on multi-scale \textit{code-switch protein sequences} and (\romannumeral2) describing relationships among residues and atoms using a \textit{multi-scale position encoding}.

First, drawing inspiration from the concept of multilingual code-switching in machine translation \citep{yang2020csp,li2022universal},\footnote{They create sentences that switch between two or more languages to help the model learn multilingual knowledge. \citet{yang2020csp} enhance multilingual model capabilities by substituting words in the source sentence with their translations in the target language. Similarly, \citet{li2022universal} improve these abilities by replacing a source word or phrase with its counterpart in a different language and then masking the corresponding target word. Collectively, these studies demonstrate that such code-switching techniques significantly strengthen the multilingual capabilities of machine translation models.} \msESM~introduces the concept of learning multi-scale knowledge by pre-training on code-switch protein sequences. These sequences are a hybrid of sequence and structure data, derived from randomly unzipping protein residues into their constituent atoms and assigning coordinates to each unzipped atom.
In such a scenario, \msESM~can not only capture multi-scale aligned knowledge but also efficiently handle residue sequences and atomic coordinates.

Second, \msESM~employs a multi-scale position encoding to comprehensively differentiate between residues and atoms within the code-switch protein sequence.
At the residue scale, we extend the original position encoding used in ESM to align with the current best practices in handling pure residue sequences, thereby avoiding ambiguous positional information across different scales, including atom-to-atom, residue-to-residue, and residue-to-atom relationships.
At the atom scale, to describe the relationships among unzipped atoms, we employ a spatial distance matrix that directly encodes their 3D positions.
With this approach, we can effectively describe all relationships among the entities within the code-switch sequence.

We pre-train \msESM~using a mixture of protein and small molecule data, and fine-tune it on a diverse set of benchmarks for evaluation. The improved experiment results demonstrate that \msESM~surpasses previous methods in protein-molecule tasks, indicating the full utilization of protein language models.
The solid performance in protein tasks suggests that \msESM, facilitated by the novel unified molecular modeling we first proposed, acquires molecular knowledge without sacrificing its understanding of proteins.
Additionally, when applying \msESM~to standard molecular benchmarks, it also outperforms several molecule-specific models.
These findings clearly highlight the potential of unified molecular modeling.

\begin{figure*}[!t]
\centering
% \vspace{-1.0cm}
\vspace{-0.3cm} 
\subfigure[Multi-scale MLM]{\includegraphics[width=0.44\textwidth]{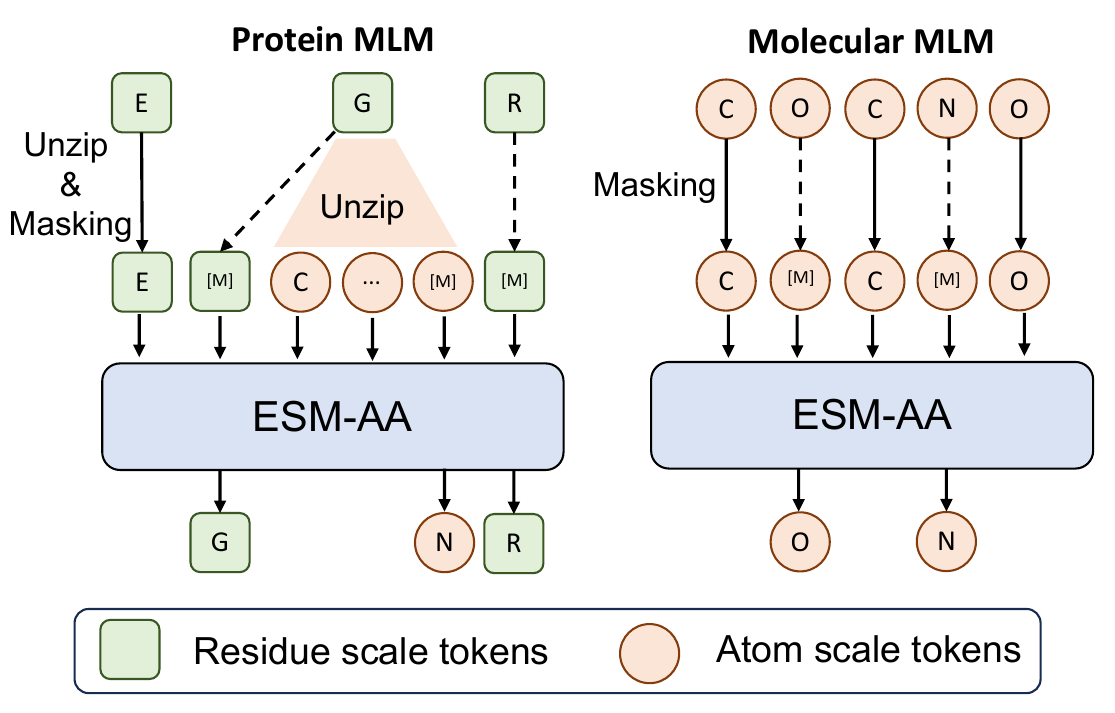}
}
\hspace{0.01\textwidth}
\subfigure[Pair-wise Distance Recovery]{\includegraphics[width=0.45\textwidth]{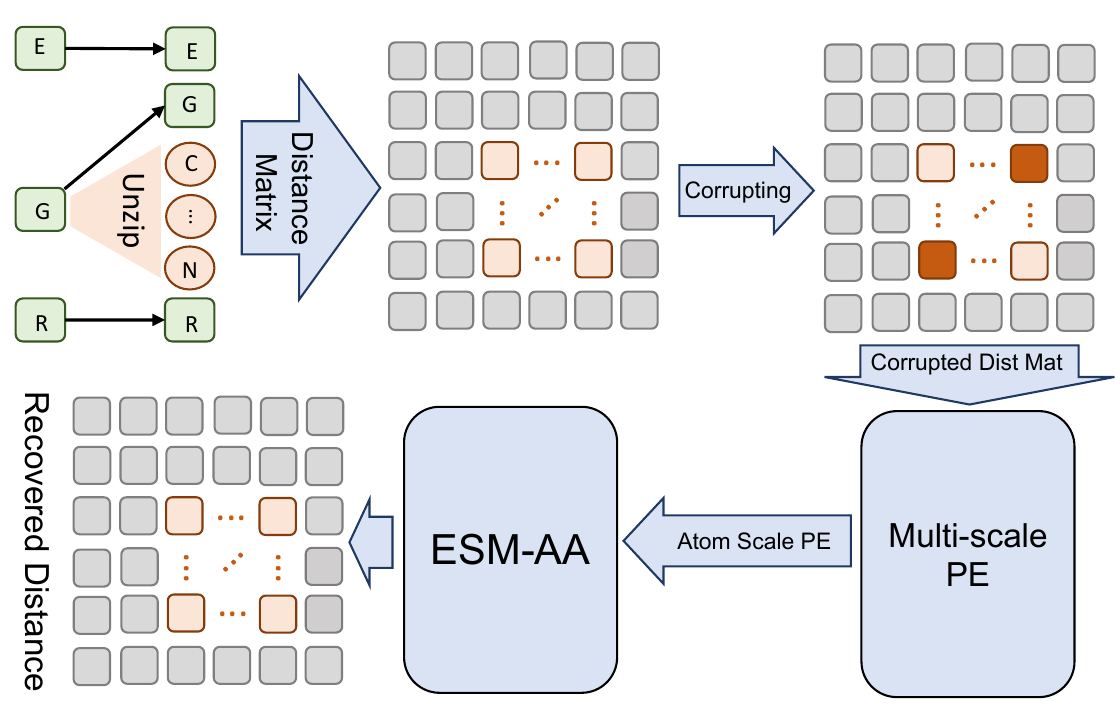}}
\caption{Framework of multi-scale pre-training comprises multi-scale masked language modeling and pairwise distance recovery.}
\label{fig_mathod}
% \vspace{-0.6cm}
\vspace{-0.3cm} 
\end{figure*}

\section{Proposed Method: \msESM}
\label{multi_scale_pre_training}
In this section, we will describe our multi-scale pre-training model, i.e., \msESM, in detail.
Due to the vast number of atoms in a protein molecule, it is impossible to simultaneously input all atomic information of a protein into the model. Inspired by the concept of multi-lingual code-switching methods, \msESM~initially generates multi-scale code-switch protein sequences by randomly unzipping partial residues.
Through training on these sequences with carefully designed multi-scale position encoding, \msESM~demonstrates its efficacy at both the residue and atom scales.
When addressing protein-molecule tasks, i.e., tasks involving both proteins and small molecules, \msESM~does not require any additional models and can fully leverage the potential of pre-training.

Specifically, in Section \ref{overview}, we introduce the overall objective of training \msESM.
Subsequently, in Section \ref{hybrid_structure}, we delve into the details of constructing a code-switch protein sequence and implementing the multi-scale pre-training approach.
To describe the complicated position relationships within the code-switch sequence, we present our design of a multi-scale position encoding in Section \ref{unified_relative_position_encoding}.

\subsection{Overview}
\label{overview}
We start with an overview of our multi-scale pre-training model, i.e., \msESM~(Figure \ref{fig_overview}).
Briefly, the total objective of our pre-training can be expressed as the following loss function:
\begin{align}
\mathcal{L}_{\theta} = \sum_{X_i \in B} &
 \mathcal{L}_{\textsc{mlm}}(\bar{X}_{i}, E_{i};\theta)+  \mathcal{L}_{\textsc{pdr}}(\bar{X}_{i}, E_{i};\theta) \nonumber \\
 =\sum_{X_i \in B} & \mathcal{L}_{\textsc{mlm}}(\textsc{Unzip}(X_i), \textsc{mspe}(X_i);\theta) ~+ \nonumber \\ & 
 \mathcal{L}_{\textsc{pdr}}(\textsc{Unzip}(X_i), \textsc{mspe}(X_i);\theta)  \nonumber
\end{align}
where $B$ is a batch of data sampled from the dataset $D$.
For each data $X_i$ in dataset $D$, we first create its code-switch sequence $\bar{X}_{i}$ by unzipping partial residues.
Using the code-switch sequence, we employ Masked Language Modeling (MLM) and Pair-wise Distance Recovery (PDR) as pre-training tasks.
We discuss the details of $\bar{X}_{i}$, $\mathcal{L}_{\textsc{MLM}}$, and $\mathcal{L}_{\textsc{PDR}}$ in Section \ref{hybrid_structure}.
To account for the coexistence of residues and atoms in the sequence, we propose a Multi-Scale Position Encoding (MSPE) $E_i$ to describe the complicated position relationship within $\bar{X}_{i}$ (Section \ref{unified_relative_position_encoding}).
We show more details of \msESM, including the parameterization of $\theta$ in Section \ref{pre_training}.
Notably, since we utilize molecule data in pre-training, \msESM~can accept either proteins or molecules as inputs.

\subsection{Multi-scale Pre-training}
\label{hybrid_structure}
In this section, we elaborate how to create a code-switch protein sequence $\bar{X}$ and implement the pre-training tasks, i.e., Masked Language Modeling (MLM) and Pair-wise Distance Recovery (PDR), on it (Figure \ref{fig_mathod}).

\paragraph{Code-Switch Protein Sequence}
Briefly, the concept of constructing a code-switch protein sequence is inspired by the multilingual code-switching technique in machine translation \citep{yang2020csp,li2022universal}.
This technique, which constructs sentences that switch between multiple languages, has significantly enhanced the model's capability to handle multilingual tasks.
In our multi-scale unified molecular modeling, we treat residues and atoms as different ``languages" and construct sequences that switch between residues and atoms, thereby augmenting the model's capability to handle downstream tasks.

Specifically, in the residue scale, a protein $X$ can be seen as a sequence of $L$ residues, i.e., $X = (r_1, \cdots, r_{i}, \cdots, r_{L})$.
Each residue $r_{i}$ further consists of a specific set of $N$ atoms $A_i = \{a_{i}^{1}, \cdots, a_{i}^{N}\}$.
To construct a code-switch protein sequence $\bar{X}$, we randomly select a group of residues and insert their corresponding atoms into $X$, which is essentially an unzipping process. For each unzipped residue, we provide the model with structural information of the residue at the atomic scale, i.e., atomic coordinates, thus offering the model very diverse structural knowledge. In particular, during the unzipping process, we assign a sequential order to the unzipped atoms.
Here, we take the case of unzipping a single residue as an example, whereas in actual modeling, multiple residues can be unzipped. 
After inserting the atom set $A_i$ into $X$, i.e., unzipping the residue $r_i$, we obtain a code-switch sequence
\begin{align*}
\bar{X} &= (r_1, \cdots, r_i, \textsc{Order}(A_i), \cdots, r_L) \\
&=(r_1, \cdots, r_{i}, a_{i}^{1}, \cdots, a_{i}^{N}, \cdots, r_L) \\
&=(h_1, \cdots, h_{i}, h_{i+1}, \cdots,  h_{i + N}, \cdots, h_{L+N})
\end{align*}
where $\textsc{Order}$ is the order assigned to the atom set (Appendix \ref{pre_training_config}).
$h_i$ represents either a single residue or an individual atom in $\bar{X}$.
We also denote all the atoms in $\bar{X}$ as $\bar{A}$ and all the residues as $\bar{R}$.

Notably, when we insert the atom set $A_i$ of residue $r_i$, we still retain $r_i$. 
This allows the model to attend either to the corresponding residue-scale information or to the surrounding atom-scale information when predicting masked atoms and encourages the model to align residue-scale and atom-scale representations, similar to the approach in cross-lingual pre-training \citep{conneau2019cross}.
We provide an illustration of the code-switch sequence in Figure \ref{fig_mathod}.

\paragraph{Masked Language Modeling}
After obtaining the code-switch sequence $\bar{X}$, we can implement MLM on it.
Unlike the MLM used in ESM, which only masks residues, our approach masks both residues and atoms and requires models to predict them.
Specifically, we start by randomly masking a portion of the atoms or residues in $\bar{X}$ and then ask the model to predict the original atoms or residues using the surrounding context.
$$
\mathcal{L}_{\theta\textsc{MLM}} = - \sum_{h \in \textsc{Mask}(\bar{X})} \log p_{\theta}(h|\bar{X}\backslash \textsc{Mask}(\bar{X}))
$$
where $\textsc{Mask}(\bar{X})$ represents the set of masked atoms and residues.
$\bar{X} \backslash \textsc{Mask}(\bar{X})$ denotes the unmasked context.
$h$ is a single masked atom or residue.
Figure \ref{fig_mathod}a is the framework of MLM task.

\paragraph{Pair-wise Distance Recovery}
We also employ PDR as another pre-training task.
Briefly, we use corrupted atoms as model input and ask model to recover the accurate Euclidean distances between these atoms.
We corrupt the atoms by adding noises to their coordinates.
Specifically, we replace the ground-truth coordinate with a randomly selected position that is within a certain range (Euclidean distances $< \epsilon$, Appendix \ref{pre_training_config}) of the true coordinate.
Models are required to reconstruct the actual distances based on the corrupted coordinates.
To avoid introducing residue-residue interactions that are very different from the interactions in small molecules, we only calculate PDR within residues, which can also make \msESM~learn very diverse structural knowledge of residues.
$$
\mathcal{L}_{\theta \textsc{PDR}} = \sum_{\substack{A_i = A_j\\h_i, h_j \in \bar{A},i \not = j   \\ c_i = \textsc{Coord}(h_i) \\ c_j = \textsc{Coord}(h_j)}} \Vert \textsc{Dis}_{\theta}(c_i + \sigma_i,c_j+\sigma_j)  - \textsc{Dis}(c_i, c_j)\Vert_{2}
$$
where $\textsc{Dis}_{\theta}$ is the recovered distance and $\textsc{Dis}$ is the ground truth.
$\textsc{Coord}$ extracts coordinates from atoms.
$\sigma_i, \sigma_j$ are the corresponding noises added to atom coordinates $c_i, c_j$.
To elaborate further, these noises will affect the atom scale position encoding in Section \ref{unified_relative_position_encoding}.
Figure \ref{fig_mathod}b shows the framework of PDR task.

Notably, when training \msESM, we mix up a protein dataset $D_{p}$ and a molecule dataset $D_{m}$ as the final dataset, i.e., $D = D_{p} \cup D_{m}$.
For a molecule from $D_{m}$, as it consists solely of atoms, its code-switch sequence $\bar{X}$ is the ordered set of all its atoms $\bar{A}$, and it does not have any residues, i.e., $\bar{R} = \emptyset$.

\begin{figure}[ht]
\centering
% \vspace{-0.9cm} 
\includegraphics[scale=0.34]{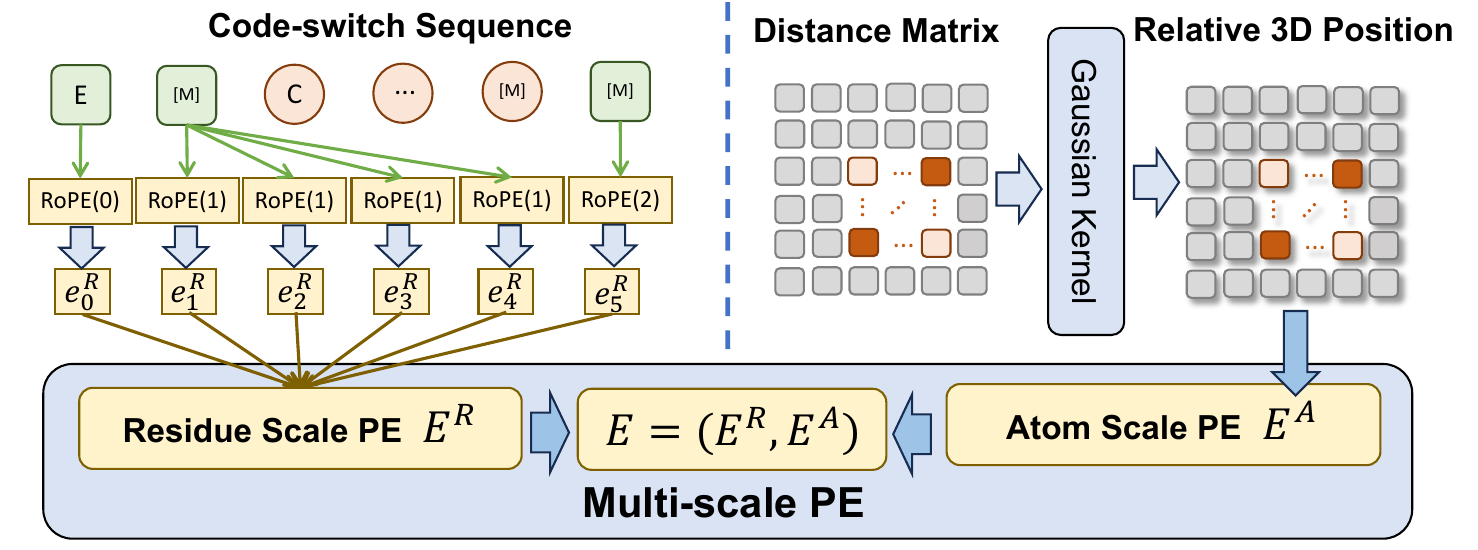}
\caption{Framework of multi-scale position encoding.}
\label{fig_pe}
\vspace{-0.3cm}
\end{figure}

\subsection{Multi-scale Position Encoding}
\label{unified_relative_position_encoding}
Encoding the position relationship in the code-switch sequence is challenging.
Given that both residues and atoms are present in the code-switch sequence, it is crucial for the position encoding to accurately represent the positional relationships.
This includes relationships between residues, between atoms, and between residues and atoms, regardless of whether the atoms are part of the same residue.
This situation is more complex than dealing with pure residue sequences.
Because previous encodings in PLMs are only designed for residue sequences, they can not describe the relationships that extend from residues to atoms, and among atoms.

In this section, we design a multi-scale position encoding $E$ to encode the positional relationships within a code-switch sequence.
Specifically, $E$ contains a residue scale position encoding $E^{R}$ and an atom scale position encoding $E^{A}$, i.e., $E = (E^{R},E^{A})$.
For $E^{R}$, we carefully extend an existing encoding method, allowing it to encode relationships from residues to atoms, while maintaining consistency with the original encoding when handling pure residue sequences.
For $E^{A}$, to capture the relationships among atoms, we directly encode their 3D positions using a spatial distance matrix.
The multi-scale encoding approach ensures that no ambiguous positional relationships affect the pre-training, enabling \msESM~to perform effectively in both scales.
Figure \ref{fig_pe} illustrates the framework of our multi-scale position encoding.
We will provide detailed explanations for each of them in the following paragraphs.

\paragraph{Residue Scale Position Encoding}
We design the residue scale position encoding $E^{R}$ following two principles: 
(\romannumeral1) For encoding the relationship between two residues, $E^{R}$ should be consistent with the mainstream encoding method. (\romannumeral2) For atoms from the same unzipped residue, $E^{R}$ should not introduce any ambiguous position information.

As previous PLMs have shown the effectiveness of the mainstream encoding method in handling pure residue sequences, it is prudent for $E^{R}$ to maintain consistency with it.
Furthermore, when dealing with two atoms from the same residue, since we cannot define residue scale positional relationships within the residue, it is important for $E^{R}$ to avoid the impact of such ill-defined information.

In particular, we use Rotary Position Embedding (RoPE) \citep{su2021roformer}, the original position encoding in ESM-2, to describe the position relationship among the residues in a code-switch sequence.
For assigning the position encoding to an atom in the code-switch sequence, we reuse the position encoding of the residue to which the atom belongs.
In cases where the atom belongs to a small molecule, not a residue, we assign a fixed position encoding (\textsc{RoPE}(0) in our paper) to it.
Formally, for a code-switch sequence $\bar{X}$, its residue scale position encoding $E^{R} = (e_1^{R}, \cdots, e_{L+N}^{R})$ can be obtained according to the following formulation:
$$
e_i^{R} = 
\left\{ \begin{array}{ll}
\textsc{RoPE}(j) & h_i \in \bar{R}, h_i = r_j \\
\textsc{RoPE}(k) & h_i \in \bar{A}, \exists k, h_i \in A_k \\
\textsc{RoPE}(0) & \text{otherwise} \\
\end{array} \right.
$$
By adopting such encoding strategy, $E^{R}$ satisfies the two aforementioned principles.
Specifically, for pure residue sequences, $E^{R}$ is equivalent to RoPE.
When handling atoms from the same residue, the relative nature of RoPE ensures that no ambiguous information will impact the pre-training model.
For more details about the properties of RoPE, please refer to \citet{su2021roformer}.

\paragraph{Atom Scale Position Encoding}
Because $E^R$ will not provide the position encoding for atoms from the same residue, we need an atom scale position encoding $E^A$ to describe the relationship from atoms to atoms.
As suggested by \citet{zhou2023uni}, we use Euclidean distance matrix and Gaussian kernel $\textsc{Gaussian}$ to encode the 3D position of atoms.

For $h_i, h_j \in \bar{X}$, their atom scale position encoding $e_{ij}^{A}$ can be calculate as follows:
$$
e_{ij}^{A} = 
\left\{ \begin{array}{ll}
0 & h_i \in \bar{R} \text{~or~} h_j \in \bar{R} \\
\textsc{Gaussian}(\textsc{Dis}(c_i, c_j)) & \text{otherwise}
% \\ & c_i = \textsc{Coord}(h_i), \\ & c_j = \textsc{Coord}(h_j)
\end{array} \right.
$$
where $c_i = \textsc{Coord}(h_i), c_j = \textsc{Coord}(h_j)$. 
We refer readers to \citet{zhou2023uni} for more details of this 3D position encoding.

% \subsection{Other details of \msESM}
\subsection{Integrating Multi-scale PE into Transformer}
\label{pre_training}
The parameterization $\theta$ of \msESM~is slightly different from the original Transformer architecture proposed by \citet{vaswani2017attention}.
To be specific, we begin by substituting the sinusoidal encoding in the Transformer with our residue scale position encoding $E^{R}$.
For the atom scale position encoding $E^{A}$, we treat it as the bias term of self-attention layers \citep{luo2022one,zhou2023uni}.
The self-attention in \msESM~can be calculated like:
$$
\textsc{Attention}(Q,K,V, E^{A}) = \textsc{softmax}(\frac{QK^{T}}{\sqrt{d_k}} + E^{A}) V
$$
where $Q, K, V$ are the query, key, and value corresponding to $\bar{X}$.
We refer readers to \citet{vaswani2017attention} for more details of the original Transformer.
With only slight modifications to the original Transformer architecture, \msESM~is capable of simultaneously processing residues and atoms, making it a versatile model for various downstream tasks.
Moreover, \msESM~shows great compatibility with existing pre-training model, e.g., ESM series, which allows us to bulid up a better model based on previous study more easily.

\begin{table*}[t]
\footnotesize
% \scriptsize
\color{black}{
\caption{Performance comparison on Enzyme-Substrate Affinity Regression (ESAR) task and Enzyme-Substrate Pair Classification (ESPC) task. \msESM~outperforms other models and achieves the
state-of-the-art results, which indicates that \msESM~operate at both the residue and atom scales successfully and our unified modeling harness the full potential of PLMs.}
\label{table_protein_molecule_esa}
\begin{center}
\begin{tabular}{c|cc|ccc|ccc}
% \vspace{-0.5cm} 
\toprule
\multicolumn{1}{c|}{\multirow{2}{*}{Method}} &\multicolumn{1}{c}{Protein} &\multicolumn{1}{c|}{Molecule} & \multicolumn{3}{c|}{\multirow{1}{*}{ESAR}} &\multicolumn{3}{c}{\multirow{1}{*}{ESPC}} \\ 
& Pre-training & Pre-training & MSE $\downarrow$ & $R^2 \uparrow$& Pearson $\uparrow$& ACC $\uparrow$& MCC $\uparrow$&ROC-AUC $\uparrow$ \\
\midrule
\citet{gollub2023enkie} & / & / & / & 0.463 & 0.680 & / & / & /\\
\citet{kroll2021deep} & / & / & 0.653 & 0.527 & 0.728& / & / & / \\
\midrule
$\text{Baseline}_{\text{ XGBoost}}$ & ESM-2 35M & Uni-Mol 48M & 0.652 & 0.528 & 0.727 & 89.9\% & 0.729 & 0.941\\
$\text{Baseline}_{\text{ ProSmith}}$ & ESM-2 35M & Uni-Mol 48M & 0.642 & 0.536 & 0.733 & 90.8\% & 0.754 & 0.943\\
\midrule
\rowcolor{mygray} $\text{Ours}_{\text{ XGBoost}}$ & \msESM~35M& \msESM~35M& 0.620 & 0.551 & 0.744 & 90.4\% & 0.743& 0.949\\
\rowcolor{mygray}$\text{Ours}_{\text{ ProSmith}}$ & \msESM~35M& \msESM~35M& {\bf 0.607} & {\bf 0.560} & {\bf 0.752} & {\bf 92.3\%} & {\bf 0.797} &  {\bf 0.957}\\
\bottomrule
\end{tabular}
\end{center}
}
\vspace{-0.2cm} 
\end{table*}

\begin{table*}[t]
\vspace{-0.4cm} 
\footnotesize
\color{black}{
\caption{Performance comparison on drug-target affinity regression task. 
\msESM~achieves the state-of-the-art results on most metrics.
}
\label{table_protein_molecule_dta}
\begin{center}
\begin{tabular}{c|cc|ccc}
\toprule
\multicolumn{1}{c|}{\multirow{2}{*}{Method}} &\multicolumn{1}{c}{Protein} &\multicolumn{1}{c|}{Molecule} &\multicolumn{1}{c}{\multirow{2}{*}{MSE $\downarrow$}} &\multicolumn{1}{c}{\multirow{2}{*}{CI $\uparrow$}} &\multicolumn{1}{c}{\multirow{2}{*}{$r^{2}_{m} \uparrow$}}\\ 
& Pre-training & Pre-training & & &\\
\midrule
\citet{ozturk2018deepdta} & / & / & 0.261  & 0.878  & 0.630 \\
\citet{shin2019self} & / & Molecule Transformer & 0.245 & 0.887 & 0.665 \\
\citet{nguyen2021graphdta} & / & / & 0.229 & 0.893 & 0.685 \\
\citet{nguyen2021gefa} & TAPE 38M & / & 0.228 & 0.893 & / \\
\citet{qiu2021rzmlp} & ProtBert 420M& / & 0.205 & 0.896 & 0.709 \\
\citet{kao2021toward} & /	&/	&0.202 & 0.907& / \\
\citet{yuan2022fusiondta}& ESM-1b 650M &/&	0.208 &	{\bf 0.913} 	&0.743 \\ 
\citet{yang2022mgraphdta}&/	&/	&	0.207 &	0.900 &	0.710 \\
\midrule
$\text{Baseline}_{\text{ XGBoost}}$ & ESM-2 35M & Uni-Mol 48M & 0.261 & 0.885 & 0.652 \\
$\text{Baseline}_{\text{ ProSmith}}$ & ESM-2 35M & Uni-Mol 48M & 0.219 & 0.899 & 0.711 \\
\midrule
\rowcolor{mygray}$\text{Ours}_{\text{ XGBoost}}$ & \msESM~35M& \msESM~35M& 0.243 & 0.890 & 0.678 \\
\rowcolor{mygray}$\text{Ours}_{\text{ ProSmith}}$ & \msESM~35M& \msESM~35M& {\bf 0.196} & 0.903 & {\bf 0.752} \\
\bottomrule
\end{tabular}
\end{center}
}
\vspace{-0.6cm} 
\end{table*}

\section{Experiments}
\label{results_discussions}

We pre-train \msESM~on mixed data of proteins and small molecules. 
For the proteins, we construct code-switch sequences that contain both sequence and structural information, as described in Section \ref{hybrid_structure}.
We fine-tune and evaluate \msESM~across diverse benchmarks and verify the contribution of each component through ablation experiments. Finally, a visualization analysis is included to explain the advantages of unified modeling. 

\subsection{Pre-training Configuration}
\paragraph{Datasets}
We pre-train using a dataset that includes both protein and molecule data, specifically selecting those with structural details such as atom coordinates for encoding Euclidean distances and recovering pair-wise distances.
For the protein dataset, we use AlphaFold DB \citep{varadi2022alphafold} dataset, which contains 8M protein sequences and structures predicted by AlphaFold2 \citep{jumper2021highly} with high confidence (pLDDT $>$ 90).
For the molecule dataset, we use the dataset provided by \citet{zhou2023uni}, which contains 19M molecules and 209M conformations generated by ETKGD \citep{riniker2015better} and Merck Molecular Force Field \citep{halgren1996merck}.

\paragraph{Hyperparameters}
We implement \msESM~using 12 stacked Transformer layers, each with 20 attention heads, as discussed in Section \ref{pre_training}.
The model dimension and feedforward dimension of each Transformer layer are 480 and 1920.
We use Adam \citep{kingma2014adam} and polynomial learning rate scheduler to train \msESM~and set the learning rate 4e-4, weight decay 1e-2, warmup step 5000.
The total training step is 300K and each batch has 256K tokens at maximum.
We train \msESM~on 16 NVIDIA A100 GPU cards for 3 days.
\msESM~is compatible with ESM series, so we load a ESM-2 checkpoint as the initialization of \msESM.
When pre-training, 1.0\% of residues are unzipped as the pre-training setting, which makes the unzipped protein sequence 1.08 times longer than before on average. Thus we make an adjustment to the maximum sequence length permissible for \msESM, transitioning from ESM-2's 1024 to 2048.
Table \ref{table_hyper} provides a complete list of hyperparameters.

\subsection{Main Results}
\label{protein_molecule_tasks}
We use tasks involving both proteins and molecules to prove that \msESM~can operate at both residue and atom scales and our unified molecular modeling approach can exploit the full potential of PLMs.

\paragraph{Fine-tuning} 
For protein-molecule tasks, we follow the benchmark protocol from ProSmith \citep{kroll2023multimodal} to evaluate \msESM~on three tasks, including enzyme-substrate affinity regression, drug-target affinity regression, and enzyme-substrate pair classification.
Specifically, each task provides the protein residue sequence and the molecule SMILES string as input and requires models to determine whether the protein-molecule pair exhibits high affinity.
Since our \msESM~cannot directly process SMILES strings, we initially employ RDKit \citep{landrum2013rdkit} to generate the corresponding molecule conformations based on the SMILES representation. 
Subsequently, we extract the atom sequence and atom scale position encoding for \msESM. 
For additional fine-tuning details (datasets and hyperparameters), please refer to Appendix \ref{md_pm_task}.

\paragraph{Baselines}
We compare \msESM~with multiple baselines on each tasks, including both supervised and pre-training baselines.
For each baseline, we list their protein pre-training model and molecule pre-training model in corresponding tables.
More details of each baseline can be seen in corresponding papers. We also use the standard framework provided by ProSmith for evaluating \msESM~to ensure a fair comparison.
Specifically, the framework contains three main modules, i.e., molecule encoder, protein encoder, and fusion block. 
Two encoders extract features from proteins and molecules severally. 
The fusion block is a Transformer model, which is responsible for fusing protein and molecule features. 
The fused features are further used to regress the affinity values or predict binary affinity.
We compare performance by replacing encoders with different pre-trained models (\msESM, ESM-2, Uni-Mol).
We also provide the results of an XGBoost \citep{chen2016xgboost} variant of ProSmith, which removes the fusion block and uses simple concatenation for feature fusing and can directly assess the compatibility of the two representations. 
Note that we freeze both encoders in the experiments as suggested by ProSmith.
We turn off the unzip operation when performing fine-tuning.

\paragraph{Results} 
Table \ref{table_protein_molecule_esa} and Table \ref{table_protein_molecule_dta} display the experimental results of \msESM~and baselines for the three tasks.
Based on the results, we can summarize our findings as follows:
(\romannumeral1) \msESM~outperforms other models and achieves the state-of-the-art results on most metrics.
(\romannumeral2) Fine-tuning strategies such as ProSmith and XGBoost, when built upon our \msESM, consistently outperform versions that combine two separate pre-training models (as shown in the last four rows of both Table \ref{table_protein_molecule_esa} and Table \ref{table_protein_molecule_dta}).
(\romannumeral3) \msESM~can even beat methods that are based on much larger pre-training models (comparing the 5th and 7th rows to the last row in Table \ref{table_protein_molecule_dta}).

These findings clearly indicate that \msESM~operate at both the residue and atom scales successfully and {\bf pre-training proteins and molecules in a single model can harness the full potential of pre-training techniques for protein-molecule tasks}.
Fusing two separate pre-training models can be suboptimal for such tasks, and the issue cannot be resolved by using larger pre-training models.

\begin{table}
\vspace{-0.5cm} 
\footnotesize
\label{table_mol}
\centering
\caption{Experimental results on ablation study. The results show that each component contributes to our method.}
\begin{tabular}{c|cc}
\toprule
\multicolumn{1}{c|}{\multirow{2}{*}{Method}} &\multicolumn{2}{c}{\multirow{1}{*}{ESAR}}  \\
 & MSE $\downarrow$ & $R^2 \uparrow$ \\
\midrule
w/o ASPE  &  0.639(+0.012) & 0.537(-0.009) \\ 
w/o RSPE  & 0.676(+0.049) & 0.511(-0.035) \\
\midrule
w/o MLM Loss  & 0.642(+0.015) & 0.535(-0.011) \\
w/o PDR Loss  & 0.645(+0.018) & 0.533(-0.013) \\
\midrule
w/o Molecule Data  & 0.648(+0.021) & 0.531(-0.015) \\
w/o Protein Data  & 0.708(+0.081) & 0.487(-0.059) \\
w/o Unzip Operation  & 0.638(+0.011) & 0.538(-0.008) \\
\midrule
\rowcolor{mygray} \msESM  & {\bf 0.627} & {\bf 0.546} \\
\bottomrule
\end{tabular}
\label{more_ablation}
\vspace{-0.5cm} 
\end{table}

\subsection{Ablation Study}
\label{ablation}
We have conducted comprehensive ablation studies focusing on various aspects such as position encoding, pre-training objectives, and training data.
These studies demonstrate that each of these components plays a crucial role in the efficacy of our method. We also provide an analysis of different pre-trained model combinations in Appendix \ref{mar}.
The results further confirm the effectiveness of the strategy for unified processing of proteins and molecules.

\paragraph{Ablation on  Multi-scale Position Encoding}
\label{ab_pe}
To validate the effectiveness of multi-scale position encoding, we conduct ablation tests under two conditions: one without using Atom Scale Position Encoding (ASPE) and another without using Residue Scale Position Encoding (RSPE).
The employed task is enzyme-substrate affinity regression.
As shown in Table \ref{more_ablation}, when atom scale position encoding or residue scale position encoding is omitted, the model's performance suffers significantly.
This is due to the model's inability to capture positional information of atoms and residues in the absence of position encoding.
These results prove the effectiveness of our multi-scale position encoding.

\paragraph{Ablation on Pre-training Objectives}
\label{ab_obj}
We observed a substantial decrease in model performance when we omitted either the masked atom type prediction loss or the pairwise distance recovery loss, as demonstrated in Table \ref{more_ablation}.
Notably, the omission of the pairwise distance recovery loss leads to a more substantial performance deterioration compared to the omission of the masked atom type prediction loss.
This is likely because, without the pairwise distance recovery loss, \msESM~cannot learn structural information at the atom scale.
These results suggest that, while both atom type and structural information are crucial for atom-scale details, structural information is of greater significance.

\paragraph{Ablation on Pre-training Data}
We observed a significant decrease in model performance when excluding either molecular or protein data, as depicted in Table \ref{more_ablation}.
It is interesting to note that removing protein data results in a more significant performance decline compared to omitting molecule data.
This suggests that when the model is not trained with protein data, it rapidly loses protein-related knowledge, leading to a notable drop in overall performance.
However, the model can still acquire atomic scale information through unzip operations even without molecule data.
Hence, the model performs better without molecule data compared to the scenario without protein data.
Furthermore, the model's performance significantly deteriorates when the unzip operation is omitted. These results confirm the effectiveness of the unzip operation.

\subsection{\msESM~Preserves the Strong Ability of Protein Understanding}
\label{msesm_protein}

\begin{table*}[h]
\vspace{-0.2cm} 
\color{black}{
\caption{Performance comparison on secondary structure prediction task.}
\label{table_ss_predict}
\begin{center}
\begin{tabular}{c|ccc|ccc}
\toprule
\multicolumn{1}{c|}{\multirow{2}{*}{Method}} & \multicolumn{3}{c|}{\multirow{1}{*}{SS3(ACC) $\uparrow$}}& \multicolumn{3}{c}{\multirow{1}{*}{SS8(ACC) $\uparrow$}} \\
 & cb513 & ts115 & casp12 & cb513& ts115& casp12\\
\midrule
TAPE 38M & 0.73 & 0.77 & 0.71 & 0.59 & 0.64& 0.59\\
ResNet 38M & 0.75 & 0.78 & 0.72 & 0.58 & 0.64& 0.58\\
ESM-2 35M & {\bf0.80} & {\bf0.82} & {\bf0.74} & {\bf0.65} & {\bf0.70}& {\bf0.61}\\
\midrule
\rowcolor{mygray}\msESM~35M& 0.79 & 0.81 & {\bf0.74} & 0.63 & 0.69& 0.60\\
\bottomrule
\end{tabular}
\end{center}
}
\vspace{-0.5cm} 
\end{table*}

\begin{table*}[h]
\color{black}{
\caption{Performance comparison on the unsupervised contact prediction task.}
\label{table_contact_predict}
\begin{center}
\begin{tabular}{c|ccc|ccc|ccc}
\toprule
\multicolumn{1}{c|}{\multirow{2}{*}{Method}} & \multicolumn{3}{c|}{\multirow{1}{*}{Short Range $\uparrow$}}& \multicolumn{3}{c|}{\multirow{1}{*}{Medium Range $\uparrow$}} & \multicolumn{3}{c}{\multirow{1}{*}{Long Range $\uparrow$}} \\
 & P@L & P@L/2 & P@L/5 & P@L& P@L/2& P@L/5&P@L & P@L/2& P@L/5\\
\midrule
TAPE 92M & 0.10 & 0.12&0.16&0.10&0.13&0.17&0.11&0.14&0.18\\
ESM-1 43M &  0.11& 0.13& 0.16& 0.12& 0.15& 0.19& 0.13& 0.17& 0.22\\
ESM-2 35M & 0.20 & 0.29 & 0.46 & 0.22 & \textbf{0.32}& \textbf{0.45}&\textbf{0.30}&\textbf{0.39 }& \textbf{0.49}\\
\midrule
\rowcolor{mygray}\msESM~35M& \textbf{0.21} & \textbf{0.31} & {\bf0.48} & \textbf{0.23} & \textbf{0.32}& \textbf{0.45}& 0.29&0.38 & 0.48\\
\bottomrule
\end{tabular}
\end{center}
}
\vspace{-0.5cm} 
\end{table*}
Because \msESM~is developed based on existing PLMs, we would like to determine whether it still preserves a thorough understanding of proteins.
Specifically, we follow TAPE \citep{rao2019evaluating}, ESM\citep{rao2020transformer} and use the tasks secondary structure prediction and unsupervised contact prediction to test the ability of protein pre-training models in protein structure understanding.
For secondary structure prediction, models must grasp the local protein structure, such as helices and strands.
For unsupervised contact prediction, models need a comprehensive understanding of proteins at a global level.
Notably, both \msESM~and baseline methods have exactly the same input (pure residue sequence) for these two tasks.
For more details of the fine-tuning and baselines (datasets, framework, and hyperparameters), readers can find them in Appendix \ref{md_p_task}.

We report the results of secondary structure prediction and unsupervised contact prediction in Table \ref{table_ss_predict} and Table \ref{table_contact_predict}.
While \msESM~may not achieve the best performance among the compared methods, the tables demonstrate that it performs similarly to ESM-2 in both secondary structure prediction and contact prediction. 
This indicates that {\bf \msESM~does not sacrifice its understanding of proteins}.
Promisingly, \msESM~can achieve improved protein understanding by initializing its parameters with a larger ESM-2. 

\begin{figure*}[t] 
\vspace{-0.45cm} 
\centering
\subfigure[Enzyme-substrate Pair Classification]{\includegraphics[width=0.36\textwidth]{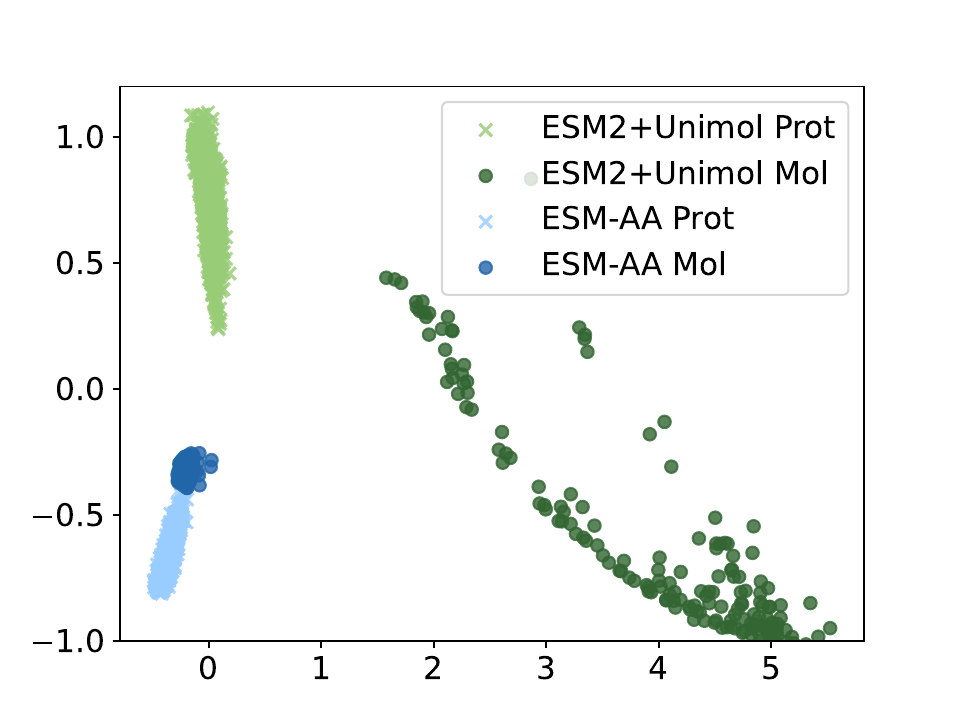}}
\hspace{0.01\textwidth}
\subfigure[Drug-target Affinity Regression]{\includegraphics[width=0.36\textwidth]{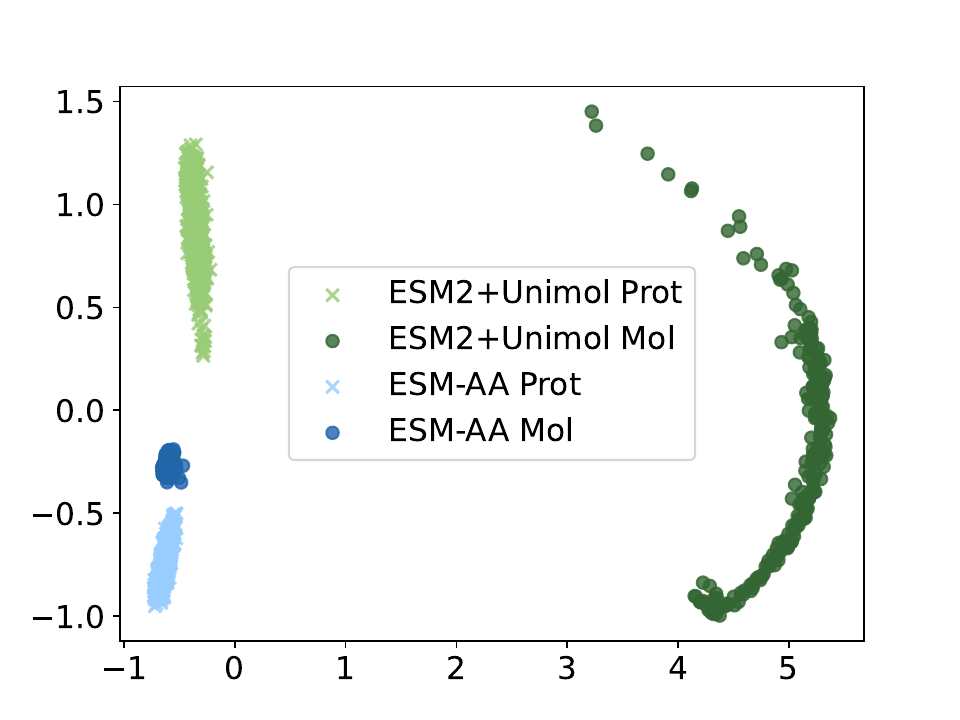}}
\caption{Visualization of representations learned by \msESM~and ESM-2+Uni-Mol.}
\label{fig_vis_rep}
\vspace{-0.3cm} 
\end{figure*}

\subsection{\msESM~Performs Well on Molecular Benchmarks}
\label{msesm_molecule}
We employ molecular benchmarks to evaluate the integrated molecular knowledge within \msESM.
Following Uni-Mol \citep{zhou2023uni}, we utilize the standard molecular benchmarks, MoleculeNet \citep{wu2018moleculenet}, in this paper.
For additional details on fine-tuning (datasets, framework, and hyperparameters) and baseline information, please refer to Appendix \ref{md_m_task}.

Table \ref{table_molecule_property} in Appendix \ref{more_exp_results} shows the experiment results of both molecular property classification and regression tasks.
\msESM~is comparable to the Uni-Mol in most tasks and outperforms several molecule-specific models in many instances, which makes it a strong method for molecular tasks.

\subsection{Visualization}
\label{vis}
To provide a more intuitive illustration of the higher quality of protein and small molecule representations learned by \msESM, we conducted a visual comparison of the representations extracted from \msESM~and ESM-2+Uni-Mol in the tasks of enzyme-substrate pair classification and drug-target affinity regression.
Specifically, we use the fine-tuned models, i.e., $\text{Baseline}_{\text{ ProSmith}}$ and $\text{Ours}_{\text{ ProSmith}}$ in both Table \ref{table_protein_molecule_esa} and Table \ref{table_protein_molecule_dta}, to extract the representations of proteins and molecules.
Subsequently, we employ Principal Component Analysis (PCA) to visualize these representations.

As illustrated in Figure \ref{fig_vis_rep}, the representations of proteins and molecules learned by the \msESM~model are more closely aligned.
This suggests that the \msESM~model is capable of creating a more cohesive semantic representation encompassing both proteins and molecular data, which makes \msESM~outperform two separate pre-trained models.

\section{Related Work}
\paragraph{Protein Pre-training}
\label{protein_pretraining}
Pre-training has been proved to be an efficient technique in many domains, like natural language processing and protein engineering.
Existing work studies protein pre-training mainly in two ways: (\romannumeral1) Sequence-based methods learn protein primary sequences to capture the biochemical and co-evolutionary knowledge.
ESM series models \citep{rives2021biological,lin2022language,lin2023evolutionary} use vanilla masked language modeling to learn protein representations on evolutionary scale.
Aiming at the specific contact prediction task, \citet{rao2021msa} further extends the masked language modeling to multiple sequence alignment (MSA) data.
Inspired by the large language model (LLM), ProtGPT2 \citep{ferruz2022protgpt2}, ProGen\citep{madani2023large}, and ProGen2 \citep{nijkamp2022progen2} scale up the model size of protein language model and show promising results in protein generation tasks.
(\romannumeral2) Structure-based methods directly learn protein structure in different levels. \citet{gligorijevic2021structure,zhang2022protein,xu2022eurnet} learn residues from a local part of protein structures.
\citet{jing2020learning,zhang2023physics} try to capture atomic structure knowledge in proteins.
We develop \msESM~based on ESM.
Differently, \msESM~is a mixture of sequence and structure-based methods, which gives it the ability to process information from different scales and makes it a versatile model.

\paragraph{Unified Molecular Modeling}
\label{multi_scale_pretraining}
Because of the huge scale difference between proteins and small molecules, it is challenging to model both of them in a unified style.
As far as we know, Uni-Mol \citep{zhou2023uni} is the only method that tries to process proteins and molecules uniformly.
Uni-Mol realizes the uniformity by directly modeling proteins and molecules at atom scale.
However, because an entire protein contains hundreds of thousands of atoms, Uni-Mol can only model a local structure of proteins, i.e., protein pocket.
Unlike Uni-Mol, as \msESM~only unzips partial residues into their corresponding atoms, it can handle an entire protein efficiently. 
Recently, GET \citep{kong2023generalist} has also considered multi-scale information for unified molecular modeling. Specifically, GET utilizes an equivariant bi-level attention module to capture residue and atom features from structures. However, GET's training strategy follows the paradigm of supervised learning, whereas \msESM~employs a method of pre-training followed by fine-tuning. We also provide some discussion of general molecular modeling in Appendix \ref{rw_mm}.

\section{Conclusions}
\label{conclusions}
In this study, we propose a multi-scale protein language model \msESM, which realizes multi-scale unified molecular modeling by pre-training on multi-scale code-switch protein sequence and describing relationships among residues and atoms with a multi-scale position encoding.
Experiment results show that \msESM~outperforms previous methods in protein-molecule tasks and effectively integrates molecular knowledge into the protein language model without sacrificing the understanding of proteins.

\section*{Acknowledgements}
We would like to thank Qiying Yu and Hanlin Wu from AIR for their insightful discussions on the project.
We also thank other members from AIR for their valuable feedback given during the internal seminar. This work is supported by the National Science and Technology Major Project (2022ZD0117502), the
National Natural Science Foundation of China (Grant No. 62276002), Natural Science Foundation of China (Grant No. 62376133) and PharMolix Inc.

\section*{Impact Statement}
PLMs have been applied to a wide range of applications, including protein structure prediction, protein fitness prediction, and protein design.
Our unified molecular modeling extends the capabilities of PLMs to effectively operate at both the residue and atom scales, thereby enhancing their applicability to these tasks.
For instance, our method can serve as the foundation for constructing more accurate protein structure prediction and design models at the atomic level.
In addition, our unified molecular modeling has also opened up new avenues for research in the field of protein-small molecule interactions.
Novel binding and drug design models can benefit from our method.
We also admit that our method inherits the potential negative influence of PLMs.
For example, it could be used to design and manufacture proteins and molecules with biological harm.

\bibliography{example_paper}
\bibliographystyle{icml2024}

%%%%%%%%%%%%%%%%%%%%%%%%%%%%%%%%%%%%%%%%%%%%%%%%%%%%%%%%%%%%%%%%%%%%%%%%%%%%%%%
%%%%%%%%%%%%%%%%%%%%%%%%%%%%%%%%%%%%%%%%%%%%%%%%%%%%%%%%%%%%%%%%%%%%%%%%%%%%%%%
% APPENDIX
%%%%%%%%%%%%%%%%%%%%%%%%%%%%%%%%%%%%%%%%%%%%%%%%%%%%%%%%%%%%%%%%%%%%%%%%%%%%%%%
%%%%%%%%%%%%%%%%%%%%%%%%%%%%%%%%%%%%%%%%%%%%%%%%%%%%%%%%%%%%%%%%%%%%%%%%%%%%%%%
\newpage
\appendix
\onecolumn
\section{Pre-training Configuration}
% \subsection{Pre-training Configuration}
\label{pre_training_config}
\paragraph{Pre-training Datasets}
We use a combined dataset consisting of both protein and molecule data for pre-training.
Since Euclidean distance is necessary for atom scale position encoding and pair-wise distance recovery, we utilize datasets that come with structural information, i.e., atom coordinates.
For the protein dataset, we use AlphaFold DB \citep{varadi2022alphafold} dataset, which contains 8M protein sequences and structures predicted by AlphaFold2 \citep{jumper2021highly} with high confidence.
For the molecule dataset, we use the dataset provided by \citet{zhou2023uni}, which contains 19M molecules and 209M conformations generated by ETKGD \citep{riniker2015better} and Merck Molecular Force Field \citep{halgren1996merck}.
Unlike \citet{zhou2023uni}, we do not train two models using two datasets respectively, instead we mix these two datasets and only train one \msESM.

\paragraph{\textsc{Order} Procedure}
For \textsc{Order} procedure, we use the default order in PDB (protein) and SDF (molecule) files as the order assigned to the atom set.
To elaborate, PDB and SDF serve as standard formats for describing atomic structures of proteins and small molecules, respectively.
In both formats, atoms follow specific sorting principles.
In our study, we directly utilize the sorted atoms for ease of implementation.
It is important to note that, given our atom scale position encoding employs Euclidean distance to describe positional relationships, the permutation of atom order does not impact our pre-training model.

\paragraph{Hyperparameters}
We implement \msESM~using 12 stacked Transformer layers, each with 20 attention heads, as discussed in Section \ref{pre_training}.
The model dimension and feedforward dimension of each Transformer layer are 480 and 1920.
The total number of \msESM's parameters is 35M.
We use Adam \citep{kingma2014adam} and polynomial learning rate scheduler to train \msESM~and set the learning rate 4e-4, weight decay 1e-2, warmup step 5000.
The total training step is 300K and each batch has 256K tokens at maximum.
We train \msESM~on 16 NVIDIA A100 GPU cards for 3 days.
\msESM~is compatible with ESM series, so we load a ESM-2 35M checkpoint as the initialization of \msESM.
When pre-training, 1.0\% of residues are unzipped as the main experimental setting, which makes the unzipped protein sequence 1.08 times longer than before on average. Thus we make an adjustment to the maximum sequence length permissible for \msESM, transitioning from ESM-2's 1024 to 2048.
For more pre-training hyperparameters, please refer to Table \ref{table_hyper}.

\begin{table}[h]
\footnotesize
\color{black}{
\caption{\msESM~ hyperparameters for pre-training.}
\label{table_hyper}

\begin{center}
\begin{tabular}{ccc}
\toprule
\multicolumn{1}{c}{\multirow{1}{*}{hyperparameters}} &\multicolumn{1}{c}{\multirow{1}{*}{Value}}\\
\midrule
Learning rate  & 4e-4 \\
LR scheduler  & polynomial\_decay \\
End learning rate & 4e-5 \\
Warmup updates & 5000 \\
Max update & 300000 \\
Max tokens & 262144 \\
Distance loss function and its weight & Smooth L1, 10.0 \\
MLM loss function and its weight & Cross entropy, 4.0 \\
Dropout & 0.0 \\
Attention dropout & 0.0 \\
Activation dropout & 0.0 \\
Num of encoder layers & 12 \\
Num of encoder attention heads & 20 \\
Encoder embedding dim & 480 \\
Encoder feedForward dim & 1920 \\
Adam ($\beta_1, \beta_2$) & (0.9,0.98) \\
Mask ratio &  0.15 \\
Unzip ratio & 0.01 \\
Distance noise $\epsilon$ & 1 \AA \\
\bottomrule
\end{tabular}
\end{center}
}
\vspace{-0.5cm} 
\end{table}

\section{Fine-tuning Details}
Here, we offer additional implementation details for fine-tuning in downstream tasks. 
We also include the statistics of each fine-tuning dataset in Table \ref{fine_tune_dataset}.

\begin{table*}
\centering
\tiny
\caption{The statistics of downstream datasets in one table. ESAR: Enzyme-Substrate Affinity Regression, DTAR: Drug-Target Affinity Regression, ESPC: Enzyme-Substrate Pair Classification, SSP: Secondary Structure Prediction, UCP: Unsupervised Contact Prediction, MPR: Molecular Property Regression, MPC: Molecular Property Classification.}
~
\label{fine_tune_dataset}
\begin{tabular}{c| c c c | c c|c c c c c c c c c c}
\toprule
\multicolumn{1}{c}{\multirow{1}{*}{}} & \multicolumn{3}{c}{\multirow{1}{*}{Protein-Molecule Task}} & \multicolumn{2}{c}{\multirow{1}{*}{Protein Task}} & \multicolumn{10}{c}{\multirow{1}{*}{Molecule Task}}\\
\midrule
\multicolumn{1}{c|}{\multirow{1}{*}{Task}} & ESAR & DTAR & ESPC & SSP & UCP & \multicolumn{3}{c}{\multirow{1}{*}{MPR}} & \multicolumn{7}{c}{\multirow{1}{*}{MPC}} \\
\midrule
\multicolumn{1}{c|}{\multirow{2}{*}{Dataset}} & \multicolumn{1}{c}{\multirow{2}{*}{KM}} & \multicolumn{1}{c}{\multirow{2}{*}{Davis}}& \multicolumn{1}{c|}{\multirow{2}{*}{ESP}} & NetSurfP‐2.0, CB513 & \multicolumn{1}{c|}{\multirow{2}{*}{ProteinNet}}  & \multicolumn{1}{c}{\multirow{2}{*}{QM7}}& \multicolumn{1}{c}{\multirow{2}{*}{QM8}} & \multicolumn{1}{c}{\multirow{2}{*}{QM9}} & \multicolumn{1}{c}{\multirow{2}{*}{HIV}} & \multicolumn{1}{c}{\multirow{2}{*}{MUV}} & \multicolumn{1}{c}{\multirow{2}{*}{BACE}} & \multicolumn{1}{c}{\multirow{2}{*}{BBBP}} & \multicolumn{1}{c}{\multirow{2}{*}{TOX21}} & \multicolumn{1}{c}{\multirow{2}{*}{PCBA}} & \multicolumn{1}{c}{\multirow{2}{*}{SIDER}}\\
 & &  & & CASP12, TS115 &  &  &  &  & & & & & & &\\
 \midrule
 Train & 8407 & 24045 & 49876 & 8678 & 20 & 5464 &17428 & 107108 & 32901 & 74469 & 1210 & 1631 & 6,264 &350343& 1141\\
 Valid & 934 & 3006 & 5540 & 2170 & 24 & 685 &2179 & 13388 & 4113 & 9309 & 151 & 204 & 783&43793& 143\\
 Test & 2335 & 3005 & 13336 & 513/21/115 & 13945 & 681 & 2179 & 13389 & 4113 & 9309 & 151 & 204 & 783&43793& 143\\
 \midrule
 Total & 11676 & 30056 & 68754 & 11497 & 13989 & 6830 &21786 & 133885 & 41127 & 93087 & 1512 & 2039 & 7830 &437929&1427\\
\bottomrule
\end{tabular}
\end{table*}
% \end{sidewaystable}

\subsection{Fine-tuning Details of Protein-Molecule Tasks}\label{md_pm_task}

\paragraph{Fine-tuning Datasets}
Following ProSmith \citep{kroll2023multimodal}, we fine-tune \msESM~and all baseline models on dataset KM \citep{kroll2021deep}, Davis \citep{davis2011comprehensive}, and ESP \citep{kroll2023general} for enzyme-substrate affinity regression, drug-target affinity regression, and enzyme-substrate pair classification respectively.
The KM dataset contains experimental affinity constants of 11676 enzyme-substrate pairs.
The Davis dataset provides 30056 binding affinities for pairs of 72 drugs and 442 proteins. The ESP dataset consists of 68754 positive or negative enzyme-substrate pairs with experimental evidence. 
We use the standard data split provided by ProSmith in fine-tuning.

\paragraph{Fine-tuning Framework}
As mentioned in Section \ref{protein_molecule_tasks}, we use ProSmith's framework for a fair comparison. 
Specifically, the framework contains three main modules, i.e., molecule encoder, protein encoder, and fusion block. 
Two encoders extract features from proteins and molecules severally. 
The fusion block is a Transformer model, which is responsible for fusing protein and molecule features. 
The fused features are further used to regress the affinity values or predict binary affinity. 
We apply our model to ProSmith's framework by replacing both protein and molecule encoders with \msESM. 
We also provide the results of an XGBoost \citep{chen2016xgboost} variant of ProSmith, which removes the fusion block and uses simple concatenation for feature fusing. 
Note that we freeze both encoders in the experiments as suggested by ProSmith.
We turn off the unzip operation when performing fine-tuning.

\paragraph{Fine-tuning Hyperparameters}
We directly use the hyperparameters provided by ProSmith. 
Specifically, the fusion block for three tasks has 6 layers of Transformer whose hidden size is 768. 
The epoch number is 100 and the learning rate is 1e-5. 
The batch sizes of the three tasks are 12, 12, and 24. 
We use Adam \citep{kingma2014adam}  as the optimizer for ProSmith and GBDT \citep{ke2017lightgbm} with 500 iterations as the predictors for XGBoost.

\subsection{Fine-tuning Details of Protein Tasks} \label{md_p_task}
\paragraph{Fine-tuning Datasets}
Following TAPE's protocol \citep{rao2019evaluating}, we evaluate \msESM~on secondary structure prediction. Specifically, for secondary structure prediction, we use data from \citet{klausen2019netsurfp} as training and validation sets and use CB513 \citep{cuff1999evaluation}, CASP12 \citep{moult2018critical}, and TS115 \citep{yang2018sixty} as test sets. The training and validation sets are filtered at the 25\% sequence identity threshold with these test tests. The final training, validation and three test sets have 8678, 2170, 513, 21, 115 protein sequences, respectively. Following ESM's protocol \citep{rao2020transformer}, we use training, validation, and test sets from ProteinNet \citep{alquraishi2019proteinnet} with training and validation sets filtered at the 30\% sequence identity threshold for unsupervised contact prediction tasks. For a fair comparison, we also remove the test data that appears in the pre-training data, and the proportion of this part of the data is less than 4‰. The final training, validation, and test sets have 20, 24, 13945 protein sequences. 

\paragraph{Fine-tuning Framework}
As suggested by TAPE, for both protein-only tasks, we use \msESM~as the protein encoder. When doing secondary structure prediction, we use a linear output layer to predict the secondary structure which each residue belongs to. When handling the unsupervised contact prediction task, we use the attention from each layer and head are independently symmetrized and corrected with APC \citep{dunn2008mutual} as features and then use a linear layer to predict whether these two residues have contact or not. Notably, both input of these two tasks is only protein sequences without structural information. Therefore, when using \msESM~to handle these two tasks, we turn off the unzip.

\paragraph{Fine-tuning Hyperparameters}
We set up all the hyperparameters aligned to TAPE. For secondary structure prediction, the epoch is 5000, batch size is 10, and learning rate is 0.001. For contact prediction, the epoch is 5, batch size 64, and learning rate is 3e-5. We use AdamW \citep{loshchilov2017decoupled} as the optimizer in secondary structure prediction and Adam \citep{kingma2014adam} in contact prediction.

\paragraph{Baselines} 
For protein tasks, we chose several popular protein pre-training models as our baselines. 
TAPE \citep{rao2019evaluating} and ResNet \citep{rao2019evaluating} employ a Transformer \citep{vaswani2017attention} and a dilated residual network \citep{yu2017dilated}, respectively, as the backbone network for training a masked language model (MLM).
Because \msESM~initializes its parameters by loading a checkpoint from ESM-2, we also include the ESM-2 model \citep{lin2023evolutionary} in our comparison.

\subsection{Fine-tuning Details of Molecule Tasks}\label{md_m_task}

\paragraph{Fine-tuning Datasets}
We use the fine-tuning data of Uni-Mol \citep{zhou2023uni} to evaluate the molecule understanding ability of \msESM. Specifically, we use QM7, QM8, and QM9 datasets for molecular property regression and HIV, MUV, BACE, BBBP, TOX21, PCBA, and SIDER datasets for molecular property classification, which have 6830, 21786, 133885, 41127, 93087, 1512, 2039, 7830, 437929, and 1427 molecules, respectively. The data split is also provided by Uni-Mol.

\paragraph{Fine-tuning Framework}
Following Uni-Mol, a special token, i.e., \texttt{[CLS]}, also exists in \msESM. Similar to NLP/CV, we simply use the representation of \texttt{[CLS]} to represent the whole molecule, and then use a linear head for fine-tuning on downstream tasks. For each molecule, we use the 3D conformation provided by \citet{zhou2023uni} as the input of \msESM. In the fine-tuning stage, we do not add noises to atom coordinates.

\paragraph{Fine-tuning Hyperparameters}
For a fair comparison, we did not search the best hyperparameters. Instead, we set up all the hyperparameters aligned to Uni-Mol. Specifically, the batch sizes for these tasks are 32, 32, 128, 256, 128, 64, 128, 128, 128, and 32. The learning rates are 3e-4, 1e-4, 1e-4, 5e-5, 2e-5, 1e-4, 4e-4, 1e-4, 1e-4, and 5e-4. The training epochs are 100, 40, 40, 5, 40, 60, 40, 80, 20, and 80. We use Adam optimizer for all these tasks.

\paragraph{Baselines} 
Following Uni-Mol, we use multiple supervised and pre-training methods as our baselines.
The details of each baseline model can be found in the Uni-Mol paper \citep{zhou2023uni}.
For a fair comparison, we evaluate the performance of the official Uni-Mol checkpoint, which uses the same molecule training data as \msESM~(remove all hydrogen atoms during training). 

\section{More Experiment Results on Molecular Tasks}
\label{more_exp_results}
Table \ref{table_molecule_property} shows the experiment results of both molecular property classification and regression tasks.

\begin{table*}[h]
\caption{Experimental results on molecular tasks. 
Compared with the vast majority of baseline models, \msESM~performs well, which demonstrates that through the unified modeling approach we enable PLMs to perform well on pure molecule tasks as well.}
\label{table_molecule_property}
\centering
\begin{tabular}{c|ccc|ccccccc}
\toprule
\multicolumn{1}{c|}{\multirow{2}{*}{Method}} & \multicolumn{3}{c|}{Reg. (MAE) $\downarrow$} & \multicolumn{7}{c}{Cls. (AUC,\%) $\uparrow$} \\
 & QM7 & QM8 & QM9 & BACE & BBBP & TOX21 & PCBA & SIDER & HIV & MUV \\
\midrule
D-MPNN & 103.5 & 0.0190 & 0.00814 & 80.9 & 71.0 & 75.9 & 86.2 &  57.0 & 77.1 & 78.6 \\
Attentive FP & 72.0 & 0.0179 & 0.00812 & 78.4 & 64.3 & 76.1 & 80.1 & 60.6 & 75.7 & 76.6 \\
$\text{N-Gram}_{\text{RF}}$ & 92.8 & 0.0236 & 0.01037 & 77.9 & 69.7 & 74.3 & - & 66.8 & 77.2 & 76.9 \\
$\text{N-Gram}_{\text{XBG}}$ & 81.9 & 0.0215 & 0.00964 & 79.1 & 69.1 & 75.8 & - & 65.5 & 78.7 & 74.8 \\
$\text{GROVER}_{\text{base}}$ & 94.5 &  0.0218 &  0.00984 & 82.6 & 70.0 & 74.3 & 76.5 & 64.8 & 62.5 & 67.3 \\
$\text{GROVER}_{\text{large}}$  & 92.0 & 0.0224 &  0.00986 & 81.0 & 69.5 & 73.5 & 83.0 & 65.4 & 68.2 & 67.3 \\
PretrainGNN & 113.2 & 0.0200 & 0.00922 & 84.5 & 68.7 & 78.1 & 86.0 & 62.7 & 79.9 & 81.3 \\
GraphMVP & - & - & - & 81.2 &72.4 & 75.9 & - & 63.9 & 77.0 & 77.7 \\
MolCLR & 66.8 & 0.0178 & - & 82.4 & 72.2 & 75.0 & - & 58.9 & 78.1 & 79.6 \\
$\text{Uni-Mol}$ & 58.9 & 0.0160 & 0.00540 & 83.2 & 71.5 & 78.9 &88.1 & 57.7 & 78.3 & 72.0 \\
\midrule
\rowcolor{mygray}$\text{\msESM}$ & 60.9 & 0.0171 & 0.00590 & 83.5 & 70.2 & 75.4 & 87.3 & 63.6 & 77.3 & 76.2 \\
\bottomrule
\end{tabular}
\end{table*}

\section{More Related Work}\label{rw_mm}

\paragraph{Molecular Modeling}
Regarding the modality of molecules, studies on molecular modeling can be categorized into three groups. 
(\romannumeral1) 1D-based methods: These represent molecules with SMILES strings and employ language modeling techniques, such as masking and contrastive self-supervision, to enhance molecular representation \citep{wang2019smiles,honda2019smiles,chithrananda2020chemberta,zhang2021mg,xue2020x,guo2022multilingual}. 
(\romannumeral2) 2D-based methods: These represent molecules with molecular graphs, sharing common ideas with general graph modeling.
Some methods \citep{rong2020self,li2020learn,zhang2021motif,li2021effective, ju2023few} mask key substructures of molecular graphs, like motifs and functional groups, and task models with reconstructing the masked parts. 
Others \citep{wang2022improving,fang2022molecular,lin2022pangu} align views from positive pairs (corrupt versions of the same graph) and simultaneously contrast views from negative pairs (different graphs).
(\romannumeral3) 3D-based methods: These directly utilize the 3D structure of molecules, aligning closely with our work. Earlier studies incorporated 3D information as an auxiliary input for 2D-based methods \citep{liu2021pre,li2022geomgcl,zhu2022unified,stark20223d}.
More recent methods focus on molecular modeling with pure 3D inputs \citep{fang2022geometry,zhou2023uni,luo2022one,zaidi2022pre,liu2022molecular,jiao2023energy}.
Three self-supervised techniques have been designed: geometry masking, geometry predicting, and denoising.
For masking, \citet{fang2022geometry} mask bond information, while \citet{zhou2023uni} mask atom types, requiring models to predict masked information based on remaining context.
For predicting, \citet{fang2022geometry} proposes an atomic prediction task with bond information to capture global structure from local information.
For denoising, models reconstruct 3D structures by adjusting corrupted structures.
When corrupting structures, \citet{zhou2023uni,luo2022one,zaidi2022pre} add Gaussian noise to each atom of the input molecule.
Several methods further introduce E(3)- and SE(3)-invariance inductive bias to the denoising technique \citep{zhou2023uni,liu2022molecular,jiao2023energy}.

\section{Performance on the Virtual Screening Benchmarks}\label{exp_vs}

We conduct pre-training experiments on inter-molecule interactions and achieved strong performance in the virtual screening benchmarks. Table \ref{table::table_vs_result} showcases the performance of models on the DUD-E zero-shot setting. The results for the baseline methods are sourced from the DrugCLIP paper\citep{gao2024drugclip}. As for DrugCLIP itself, we retrained it because the original DrugCLIP employed large-scale data augmentation, an operation we omitted during our retraining process. Based on the results presented in the table, we make the following observations:
\msESM~demonstrates robust performance, surpassing the majority of baseline methods, including widely used open-source virtual screening software Vina and commercial virtual screening software Glide-SP. This is due to \msESM's unified modeling providing a more aligned representation space for proteins and molecules, significantly enhancing the ability to screen for high-activity molecules.
Even under less-than-ideal evaluation settings, \msESM~is only slightly surpassed by the state-of-the-art, i.e., DrugCLIP. The primary reason for this is that DrugCLIP, in addition to utilizing pocket-ligand data during its secondary pre-training, also employed a significant amount of pocket data (3.2M pockets) during its initial pre-training phase. To ensure its functionality on the DUD-E benchmark, we were unable to exclude this portion of pocket data, giving DrugCLIP an unfair advantage in comparison with \msESM. However, despite its inherent disadvantage, \msESM~still achieved performance comparable to DrugCLIP, which underscores the effectiveness of its modeling strategy
\subsection{Details of the Pre-training and Finetuning} 

Following DrugCLIP\citep{gao2024drugclip}, we conducted secondary pre-training based on \msESM. This involved using protein pocket-ligand pairs as input, where pockets and ligands binding to each other served as positive samples, and randomly paired pocket-ligand combinations served as negative samples for contrastive pre-training. When processing pockets with \msESM, we decomposed each pocket residue into its constituent atoms, aligning with DrugCLIP's approach. The pre-training data, comprising over 17,000 pocket-ligand complexes from PDBBind 2019, was also sourced from DrugCLIP. Hyperparameters were largely aligned with DrugCLIP, except for the learning rate, set to 1e-4 (compared to DrugCLIP's 1e-3), as we observed that excessively high learning rates hindered \msESM~convergence.

Consistent with DrugCLIP, we also assessed the post-secondary pre-trained \msESM~using the challenging zero-shot setting from the DUD-E Benchmark, a widely recognized virtual screening benchmark. DUD-E encompasses 102 proteins and 22,886 bioactive molecules, each accompanied by 50 topologically dissimilar decoys with matched physicochemical properties retrieved from the ZINC database. To ensure the zero-shot setting, we excluded all targets present in the DUD-E from the pre-training set. We employed \msESM~to extract vector representations of both pockets and ligands, leveraging cosine similarity to rank pocket-ligand pairs, with higher cosine values indicating superior ranking. Evaluation metrics included the standard area under the receiver operating characteristic curve (AUROC), Boltzmann-enhanced discrimination of the receiver operating characteristic curve (BEDROC), and Enrichment Factor (EF).

\begin{table}[t]
\centering
\caption{Results on DUD-E in zero-shot setting. The details of baselines can be found in \citet{gao2024drugclip}.
}

\begin{tabular}{c|ccccc}
  \toprule
  Method  & AUROC(\%) $\uparrow$ & BEDROC(\%)  $\uparrow$ & EF(0.5\%)$\uparrow$ & EF(1\%)$\uparrow$ & EF(5\%)$\uparrow$\\
  \midrule
   Glide-SP&	76.70&	40.70&	19.39&	16.18&	7.23 \\
   Vina&	71.70&	-&	9.13&	7.32&	4.44 \\
   NN-score&	68.30&	12.20&	4.16&	4.02&	3.12 \\
   RFscore&	65.21&	12.41&	4.90&	4.52&	2.98 \\
   Pafnucy&	63.11&	16.50&	4.24&	3.86&	3.76 \\
   OnionNet&	59.71&	8.62&	2.84&	2.84&	2.20 \\
   DrugCLIP&	81.72&	42.24&	31.12&	26.23&	9.83 \\
  \midrule
\rowcolor{mygray} \msESM &	80.02&	39.23&	28.91&	24.12&	9.47 \\
  \bottomrule
\end{tabular}
\label{table::table_vs_result}
\end{table}

\section{Performance on Protein Function Annotation Tasks}\label{exp_pfa}

We have conducted experiments on protein function annotation tasks, where \msESM, even without structural input, matches or exceeds the performance of structural protein representation models. Table \ref{table::table_pfa_result} showcases the performance of models on the Protein Function Annotation Tasks. 

Protein Function Annotation seeks to annotate a protein with multiple functional labels. To evaluate model performance, we leverage two established benchmarks from DeepFRI\citep{gligorijevic2021structure}: Enzyme Commission (EC) number prediction and Gene Ontology (GO) term prediction. The GO benchmark further categorizes predictions into three branches: molecular function (GO-MF), biological process (GO-BP), and cellular component (GO-CC). Consistent with GearNet\citep{zhang2022protein}, we utilize the dataset splits with a 95\% sequence identity threshold for both EC and GO predictions. Notably, all models except for those explicitly defined as structural models rely solely on protein sequences as input for all tasks, including \msESM~.

\msESM~demonstrates robust performance, surpassing the majority of baseline methods. Among the selected 9 baselines, \msESM~outperforms the average performance of 8 baselines, and surpasses the ESM-2 35M model in all tasks. This demonstrates the effectiveness of our designed pretraining scheme. The \msESM~model exhibits performance close to that of GearNet, which has the highest average performance, and outperforms the average performance of other models.

\msESM~achieves or even surpasses the performance of structural models even without structural information input. The performance of \msESM~surpasses that of the protein structure model (DeepFRI, New IEConv\citep{hermosilla2022contrastive}) and approaches the performance level of the protein structure model GearNet\citep{zhang2022protein}. This indicates that even without structural information as input, \msESM~is able to model protein semantic information effectively.

\begin{table*}
\centering
\caption{Results on protein function annotation tasks. The details of baselines can be found in \citet{zhang2022protein}.
}
\label{table::table_pfa_result}

\begin{tabular}{c|cc|cc|cc|cc}
\toprule
\multicolumn{1}{c|}{\multirow{2}{*}{Method}} & \multicolumn{2}{c|}{EC}& \multicolumn{2}{c|}{GO-BP} & \multicolumn{2}{c|}{GO-MF} & \multicolumn{2}{c}{GO-CC} \\
 & AUPR & $F_{max}$ & AUPR & $F_{max}$ & AUPR & $F_{max}$ & AUPR & $F_{max}$\\
\midrule
CNN&	0.54&	0.545&	0.165&	0.244&	0.38&	0.354&	0.261&	0.387 \\
ResNet&	0.137&	0.187&	0.166&	0.28&	0.281&	0.267&	0.266&	0.403 \\
LSTM&	0.032&	0.082&	0.13&	0.248&	0.1&	0.166&	0.15&	0.32 \\
Transformer&	0.187&	0.219&	0.135&	0.257&	0.172&	0.24&	0.17&	0.38	 \\
ProtBert&	0.859&	0.838&	0.188&	0.279&	0.464&	0.456&	0.234&	0.408	 \\
DeepFRI&	0.547&	0.631&	0.282&	0.399&	0.462&	0.465&	0.363&	0.46	 \\
ESM-2 35M	& 0.803&	0.786&	0.274&	0.384&	0.582&	0.584&	0.32&	0.395	 \\
New IEConv&	0.775&	0.735&	0.273&	0.374&	0.572&	0.544&	0.316&	0.444	 \\
GearNet&	0.892&	0.874&	0.292&	0.49&	0.596&	0.654&	0.336&	0.488	 \\
\midrule
\rowcolor{mygray}\msESM~ 35M&	0.82&	0.797&	0.283&	0.401&	0.586&	0.59&	0.309&	0.418
\\
\bottomrule
\end{tabular}
\end{table*}

\section{More Ablation Results}\label{mar}
\paragraph{Ablation on Pre-trained Model Combinations} 
We further analyze the performance of different protein and molecule pre-trained model combinations on the Enzyme-Substrate Affinity Regression (ESAR) task within the framework provided by ProSmith.
The results are shown in Table \ref{tab_com}. 
Based on the data presented in the table, we make the following observations:
\begin{itemize}
    \item Utilizing a unified model to process both proteins and molecules always provides better performance than using separate models to handle each independently (last row vs. other rows). Using a unified model for proteins and molecules creates more cohesive representations of both, facilitating easier alignment of corresponding protein-molecule data for downstream tasks (as illustrated in Figure \ref{fig_vis_rep}). This approach yields better performance than employing two distinct models.
    \item Even without employing \msESM~for unified processing, using \msESM~to handle either proteins or molecules alone can also lead to performance improvements (2nd row vs. 1st row and 3rd row vs. 1st row). We believe this is due to the implicit alignment between \msESM~and both ESM-2 and Uni-Mol. Specifically, the loss function and training data used by \msESM~can be considered a combination of those from ESM-2 and Uni-Mol. Furthermore, in constructing \msESM, we also loaded the ESM-2 checkpoint for parameter initialization. This training strategy results in an implicit alignment between \msESM~and both ESM-2 and Uni-Mol, similarly offering an advantage in processing protein-molecule data.
\end{itemize}

\begin{table*}[h]

\caption{Ablation analysis of the combination of protein pre-training and molecule pre-training models. Using \msESM~for unified protein and molecule processing yields the best performance, and performance improvements are observed even when \msESM~is only used for proteins or molecules.}
\label{tab_com}
\begin{center}
\begin{tabular}{cc|ccc}
\toprule
\multicolumn{1}{c}{Protein} &\multicolumn{1}{c|}{Molecule} & \multicolumn{3}{c}{\multirow{1}{*}{ESAR}} \\ 
Pre-training & Pre-training & MSE $\downarrow$ & $R^2 \uparrow$& Pearson $\uparrow$\\
\midrule
ESM-2 35M& Uni-Mol 48M & 0.642(+0.035)& 0.536(-0.024)& 0.733(-0.019)\\
\msESM~35M& Uni-Mol 48M & 0.638(+0.031)& 0.539(-0.021)& 0.735(-0.017)\\
ESM-2 35M& \msESM~35M & 0.622(+0.015)& 0.550(-0.010)& 0.742(-0.010)\\
\midrule
\rowcolor{mygray}\msESM~35M& \msESM~35M& {\bf 0.607} & {\bf 0.560} & {\bf 0.752}\\
\bottomrule
\end{tabular}
\end{center}

\end{table*}

\paragraph{Ablation on Protein-only Tasks} 
We have tested the performance of model ablation experiments on the Contact Prediction task. And the results are shown in Table \ref{table_ablation_prot}. The absence of MLM Loss will have the most significant adverse effect on the model's performance. This is because MLM at the amino acid scale is the primary means for the model to learn semantic information about proteins. PDR (pair-wise distance recovery) performed within individual residues does not assist the model in learning global semantic information about proteins. Removing the MLM Loss will result in the model being unable to learn meaningful protein representations from the data. Removing Residue Scale Position Encoding (w/o RSPE) as well as removing protein data (w/o Protein Data) will also significantly impact the model's ability to learn protein representations. This demonstrates the necessity of Residue Scale Position Encoding. The presence or absence of the Unzip Operation does not significantly affect the model's performance on tasks such as Contact Prediction, where sequences are used as input. This indicates that the protein's local structural information introduced by the Unzip Operation does not directly impact the model's performance.

\begin{table*}[h]
\vspace{-0.3cm}
\color{black}{
\caption{The scaling experimental results on contact map prediction.}
\label{table_ablation_prot}
\begin{center}
\begin{tabular}{c|ccc|ccc|ccc}
\toprule
\multicolumn{1}{c|}{\multirow{2}{*}{Method}} & \multicolumn{3}{c|}{\multirow{1}{*}{Short Range $\uparrow$}}& \multicolumn{3}{c|}{\multirow{1}{*}{Medium Range $\uparrow$}} & \multicolumn{3}{c}{\multirow{1}{*}{Long Range $\uparrow$}} \\
 & P@L & P@L/2 & P@L/5 & P@L& P@L/2& P@L/5&P@L & P@L/2& P@L/5\\
\midrule
w/o ASPE&	0.19&	0.28&	0.43&	0.20&	0.28&	0.40&	0.25&	0.33&	0.42 \\
w/o RSPE&	0.05&	0.06&	0.05&	0.04&	0.04&	0.04&	0.02&	0.02&	0.03 \\
\midrule
w/o MLM Loss&	0.04&	0.03&	0.03&	0.03&	0.03&	0.02&	0.03&	0.03&	0.03 \\
w/o PDR Loss&	0.11&	0.14&	0.18&	0.10&	0.12&	0.15&	0.10&	0.12&	0.16 \\
\midrule
w/o Protein Data&	0.05&	0.05&	0.04&	0.04&	0.03&	0.03&	0.04&	0.04&	0.03 \\
w/o Molecule Data&	\textbf{0.20}&	0.29&	\textbf{0.46}&	\textbf{0.22}&	0.31&	\textbf{0.45}&	0.29&	0.38&	0.48 \\
w/o Unzip Operation&	\textbf{0.20}&	\textbf{0.30}&	\textbf{0.46}&	\textbf{0.22}&	\textbf{0.32}&	\textbf{0.45}&	\textbf{0.30}&	\textbf{0.38}&	\textbf{0.49} \\
\midrule
\rowcolor{mygray}\msESM~&	\textbf{0.20}&	\textbf{0.30}&	\textbf{0.46}&	\textbf{0.22}&	\textbf{0.32}&	\textbf{0.45}&	0.29&	\textbf{0.38}&	\textbf{0.48} \\
\bottomrule
\end{tabular}
\end{center}
}
\vspace{-0.3cm}
\end{table*}

\paragraph{Ablation on Molecule-only Tasks} 
Ablation studies on molecule-only tasks prove the importance of the Unzip operation in learning good molecular representations. These results are shown in Table \ref{table::table_ablation_mol}. From the results, it can be seen that the absence of ASPE and Unzip Operation will have the most significant adverse effect on the model's performance. This is because ASPE serves as the unique identifier for the model to distinguish between different atoms, while the Unzip Operation can introduce diverse residue structural information to the model. Both of these are key in improving the model's modeling of atomic-scale information. Even though the model's performance declines after removing molecular training data, it still maintains a relatively high level. This is because the Unzip operation unfolds some residues into atomic-scale information, allowing the model to learn important atomic-scale semantic representations.

\begin{table}[h]
\vspace{-0.3cm}
\centering
\caption{Ablation studies on molecule-only tasks.
}

\begin{tabular}{c|ccccc}
  \toprule
  Method  & BACE $\uparrow$ & BBBP $\uparrow$ & MUV $\uparrow$ & HIV $\uparrow$ \\
  \midrule
   w/o ASPE&	73.99&	66.41&	48.93&	75.22 \\
w/o RSPE&	75.15&	70.31&	62.62&	75.81 \\
\midrule
w/o MLM Loss&	79.15&	65.09&	65.91&	76.46 \\
w/o PDR Loss&	76.42&	64.23&	61.22&	74.55 \\
\midrule
w/o Protein Data&	74.91&	66.91&	71.49&	78.96 \\
w/o Molecule Data&	77.82&	69.12&	72.31&	72.05 \\
w/o Unzip Operation&	61.57&	65.34&	59.59&	77.21 \\
\midrule
\rowcolor{mygray}\msESM~&	\textbf{83.5}&	\textbf{70.1}&	\textbf{76.2}&	\textbf{77.3} \\
  \bottomrule
\end{tabular}
\label{table::table_ablation_mol}
\vspace{-0.3cm}
\end{table}

\section{Scaling law of \msESM}\label{scaling}

We also conducted scaling experiments at various scales and concluded that \msESM adheres to the scaling law. However, the upper limit of its capabilities on single-modal data restricts the scale-up of \msESM at this stage. Here are the detailed results.

\subsection{Scaling \msESM~ from 8M to 35M.} Scaling \msESM~ from 8M to 35M significantly improves performance on protein-molecule tasks. We test the model's performance on protein-molecule tasks(results are shown in Table \ref{tab_scaling_esar}), protein-only tasks(results are shown in Table \ref{table_scaling_prot}), and molecule-only tasks(results are shown in Table \ref{table::table_scaling_mol}), and the experimental results demonstrate that the 35M-scale \msESM~ significantly outperforms the 8M-scale \msESM~.

\subsection{Scaling \msESM~ from 35M to 150M.} Scaling \msESM~ from 35M to 150M significantly improves performance on protein-only tasks but does not result in significant performance gains for protein-molecule tasks. To further analyze the reasons behind the above phenomenon, we conducted tests on the scaled-up model's performance on molecule-only tasks. 

\par On molecule-only tasks, the performance of the 150M model showed minimal improvement. To confirm the lack of improvement in molecular performance due to scaling up, we also trained a 150M-sized Uni-Mol model. Results in Table \ref{table::table_scaling_mol} showed that the 150M Uni-Mol model exhibited almost no performance growth compared to the 47M Uni-Mol model. Further investigation into current 3D molecular representation learning models revealed that mainstream models are generally smaller than 50M, and there has been no work observed to date that extends 3D molecular representation learning models beyond 100M in size. This indicates that scaling up does not significantly benefit molecular performance for current models and datasets. On protein-only tasks, \msESM~ 150M can significantly outperform \msESM~ 35M. In summary, the bottleneck of scaling up lies in the model's ability to learn molecular representations. In the future, we will further explore how to successfully scale up molecular representation learning models. We will also incorporate these discussions and results into the paper later on.

\begin{table*}[h]
\vspace{-0.3cm}
\caption{The scaling results on enzyme-substrate affinity regression task.}
\label{tab_scaling_esar}
\begin{center}
\begin{tabular}{cc|ccc}
\toprule
\multicolumn{1}{c}{Protein} &\multicolumn{1}{c|}{Molecule} & \multicolumn{3}{c}{\multirow{1}{*}{ESAR}} \\ 
Pre-training & Pre-training & MSE $\downarrow$ & $R^2 \uparrow$& Pearson $\uparrow$\\
\midrule
ESM-2 35M& Uni-Mol 48M & 0.642& 0.536& 0.733\\
\midrule
\msESM~8M& \msESM~8M & 0.618 & 0.552 & 0.734 \\
\msESM~35M& \msESM~35M& {\bf 0.607} & {\bf 0.560} & {\bf 0.752}\\
\msESM~150M& \msESM~150M & 0.626 & 0.547 & 0.741 \\
\bottomrule
\end{tabular}
\end{center}
\vspace{-0.3cm}
\end{table*}

\begin{table*}[h]
\vspace{-0.3cm}
\centering
\caption{The scaling experimental results on molecule-only tasks.
}

\begin{tabular}{c|ccccc}
  \toprule
  Method  & BACE $\uparrow$ & BBBP $\uparrow$ & MUV $\uparrow$ & HIV $\uparrow$ \\
  \midrule
   \msESM~8M&	75.9&	67.3&	70.7&	74.3\\
   \msESM~35M&	83.5&	70.2&	\textbf{76.2}&	77.3\\
   \msESM~150M&	83.8&	71.2&	71.8&	\textbf{78.8}\\
   \midrule
   Uni-Mol 48M&	83.2&	\textbf{71.5}&	72.0&	78.3\\
   Uni-Mol 150M&	\textbf{83.9}&	71.4&	71.1&	78.5 \\
  \bottomrule
\end{tabular}
\label{table::table_scaling_mol}
\vspace{-0.3cm}
\end{table*}

\begin{table*}[t]
\vspace{-0.3cm}
\color{black}{
\caption{The scaling experimental results on contact map prediction.}
\label{table_scaling_prot}
\begin{center}
\begin{tabular}{c|ccc|ccc|ccc}
\toprule
\multicolumn{1}{c|}{\multirow{2}{*}{Method}} & \multicolumn{3}{c|}{\multirow{1}{*}{Short Range $\uparrow$}}& \multicolumn{3}{c|}{\multirow{1}{*}{Medium Range $\uparrow$}} & \multicolumn{3}{c}{\multirow{1}{*}{Long Range $\uparrow$}} \\
 & P@L & P@L/2 & P@L/5 & P@L& P@L/2& P@L/5&P@L & P@L/2& P@L/5\\
\midrule
\msESM~ 8M & 0.15&	0.21&	0.31&	0.14&	0.19&	0.26&	0.15&	0.20&	0.27\\
\msESM~ 35M & 0.21&	0.31&	0.48&	0.23&	0.32&	0.45&	0.29&	0.38&	0.48\\
\msESM~ 150M&	\textbf{0.24}&	\textbf{0.38}&	\textbf{0.60}&	\textbf{0.29}&	\textbf{0.42}&	\textbf{0.61}&	\textbf{0.43}&	0.53&	\textbf{0.69} \\
\midrule
ESM-2 35M & 0.20 & 0.29 & 0.46 & 0.22 & 0.32& 0.45&0.30&0.39 & 0.49\\
ESM-2 150M&	\textbf{0.24}&	\textbf{0.38}&	0.59	&0.28&	\textbf{0.42}	&\textbf{0.61}&	\textbf{0.43}&	\textbf{0.55}&	0.68 \\
\bottomrule
\end{tabular}
\end{center}
}
\vspace{-0.3cm}
\end{table*}

\clearpage

\section{How to Choose the Unzip Proportion}\label{unzip_ratio}

The reason for choosing an unzip proportion of 1\% is a balanced decision considering both performance and training cost. We have validated that the 1\% unzip ratio parameter is a relatively good choice through some experiments.

A high unzip ratio will incur high training costs. We find that as the unzip proportion increases, the number of tokens in the data significantly increases, as does the length of protein sequences. This leads to an increase in training cost. Therefore, choosing a proportion that is too large is not conducive to completing the training process with limited computational resourcs.
We have experimentally verified and determined the optimal unzip ratio selection.
We compared the model performance under three scenarios: unzip ratio of 0, 1\%, and 5\%. We found that when the unzip ratio is set to 1\%, the model exhibits the best performance in the protein-molecule task(shown in Table \ref{table_unzip_ratio}).
When the unzip proportion is too small, the model's performance also decreases, especially in terms of molecular representation learning performance (shown in Table \ref{table_unzip_ratio}). This is because when more residues are unfolded, the model can obtain more atomic-scale training data, which is more conducive to learning unified semantic representations.
Taking these factors into consideration, we ultimately choose 0.01 as the unzip proportion. At this proportion, approximately 8\% of the tokens in the final protein sequence are atomic-scale tokens. This falls well within our acceptable range of training costs.

% \clearpage

\begin{table*}[h]
\vspace{-0.3cm}
\color{black}{
\caption{The influence of different unzip ratios on the enzyme-substrate affinity regression task and molecule-only tasks.}
\label{table_unzip_ratio}
\begin{center}
\begin{tabular}{c|cc|ccccc}
\toprule
\multicolumn{1}{c|}{\multirow{2}{*}{Unzip Ratio}} & \multicolumn{2}{c|}{\multirow{1}{*}{ESAR}}& \multicolumn{5}{c}{\multirow{1}{*}{Molecule-only}} \\
 & MSE $\downarrow$ & $R^2 \uparrow$ & QM8$\downarrow$ & QM9$\downarrow$& HIV$\uparrow$& PCBA$\uparrow$&MUV$\uparrow$ \\
\midrule
0 &0.616&	0.554& 0.0178&	0.0068&	74.96&	86.46&	61.52 \\
1\% & \textbf{0.608}&	\textbf{0.558}& \textbf{0.0171}&	0.0059&	77.25&	87.13&	\textbf{77.3} \\
5\%& 0.618&	0.552& 	\textbf{0.0171}&	\textbf{0.0058}&	\textbf{78.05}&	\textbf{87.65}&	74 \\

\bottomrule
\end{tabular}
\end{center}
}
\end{table*}
%%%%%%%%%%%%%%%%%%%%%%%%%%%%%%%%%%%%%%%%%%%%%%%%%%%%%%%%%%%%%%%%%%%%%%%%%%%%%%%
%%%%%%%%%%%%%%%%%%%%%%%%%%%%%%%%%%%%%%%%%%%%%%%%%%%%%%%%%%%%%%%%%%%%%%%%%%%%%%%

\end{document}